\documentclass[11pt,journal,onecolumn]{IEEEtran}
\usepackage{amssymb}
\usepackage{tikz}
\usepackage{graphicx}
\usepackage[cmex10]{amsmath}
\usepackage{epsf}
\usepackage{mathrsfs}
\usepackage{cite}
\usepackage{txfonts}
\usepackage{float}
\usepackage{color}
\usepackage{textcomp}
\usepackage[usenames,dvipsnames]{pstricks}
\usepackage{epsfig}
\usepackage{pst-grad} % For gradients
\usepackage{pst-plot} % For axes
\usepackage{pst-node}
\usepackage{pst-3d}
\usepackage{etex}

\newtheorem{theorem}{\textbf{Theorem}}

\newtheorem{definition}{Definition}
\newtheorem{corollary}{Corollary}

\newtheorem{lemma}{\textbf{Lemma}}

\newtheorem{rem}{\textbf{Remark}}

\newtheorem{conj}{Conjecture A}

\newtheorem{Dirichlet2}{Theorem (Dirichlet for Vectors)}
\newtheorem{kgt}{Theorem({Khintchine-Groshev})}

\newcommand{\mb}[1]{\mathbb{#1}}
\newcommand{\mbf}[1]{\mathbf{#1}}
\newcommand{\mcal}[1]{\mathcal{#1}}
\newcommand{\hm}[1]{#1\nobreak\discretionary{}{\hbox{\ensuremath{#1}}}{}}

\def\antenna{%
\begin{pspicture}(1,1)
\pstriangle[gangle=-180.0](0,0.25)(0.6,0.25) \psline(0,0)(0,-0.60)
\psline(-0.22,-0.60)(0.22,-0.60)
\end{pspicture}
}

\begin{document}
\title{Layered Interference Alignment:\\Achieving the Total DoF of MIMO X Channels}
\author{Seyyed~Hassan~Mahboubi,~Mumtaz~Hussain,$^\dag$ ~Abolfazl~Seyed~Motahari and Amir~Keyvan~Khandani
\\\small Department of Electrical and Computer Engineering,  University of Waterloo
\\\small Waterloo, ON, Canada N2L3G1
\\ \{shmahbou,abolfazl,khandani\}@cst.uwaterloo.ca

\dag School of Mathematical and Physical Sciences, The University of Newcastle,
Callaghan, NSW 2308, Australia \\ mumtaz.hussain@newcastle.edu.au}

\maketitle
%%%%%%%%%%%%%%footnote%%%%%%%%%%%%%%%%%%%%%%%%%%%%%%%%%
\IEEEoverridecommandlockouts
\footnotetext{Financial support provided by Nortel with
the corresponding matching funds from the Natural Sciences and
Engineering Research Council of Canada (NSERC) and the Ontario Ministry
of Research and Innovation (ORF-RE) are gratefully acknowledged.\\
\dag Research is supported by the  Australian Research Council.}
\let\thefootnote

%%%%%%%%%%%%%Abstract%%%%%%%%%%%%%%%%%%%%%%%%%%%%%%%%%%%%%%
\begin{abstract}
The $K\times 2$ and $2\times K$, Multiple-Input Multiple-Output (MIMO) X channel with constant channel coefficients available at all transmitters and receivers is considered. A new alignment scheme, named \emph{layered interference alignment}, is proposed in which both vector and real interference alignment are exploited, in conjunction with joint processing at receiver sides. Data streams with fractional multiplexing gains are sent in the desired directions to align the interfering signals at receivers. To decode the intended messages at receivers, a joint processing/simultaneous decoding technique, which exploits the availability of several receive antennas, is proposed. This analysis is subsequently backed up by metrical results for systems of linear forms. In particular, for such linear forms, Khintchine--Groshev type theorems are proved over real and complex numbers.   It is observed that $K\times 2$ and $2\times K$, X channels with $M$ antennas at all transmitters/receivers enjoy duality in Degrees of Freedom (DoF). It is shown that incorporating the layered interference alignment is essential to characterize the total DoF of $\frac{2KM}{K+1}$ in the $K\times 2$ and $2\times K$, $M$ antenna X channels.
\end{abstract}

%%%%%%%%%%%%%%%%%%Keywords%%%%%%%%%%%%%%%%%%%%%%%
%\newpage
%%%%%%%%%%%%%%%%%%%% Introduction%%%%%%%%%%%%%%%%
\section{Introduction}\label{sec intro}
Sharing the available wireless medium for higher data transmission has made
interference management one of the most important challenges in wireless
networks.  However, in dense networks,
achieving the optimum throughput of the system is not necessarily obtained
by orthogonal schemes, making interference management
inevitable. Extensive efforts have been made  to characterize the
ultimate obstruction that interference imposes on the capacity of
wireless networks. In order to reduce the
severe effect of interference for the $K>2$ users interference channel,
the use of a new technique known as interference alignment is crucial.

Interference Alignment was first introduced by
Maddah-Ali et al. \cite{maddah2008communication} in the context of
Multiple-Input Multiple-Output (MIMO) X channels. It
renders the interference less damaging by merging the communication
dimensions occupied by interfering signals. Interference alignment in
$n$-dimensional Euclidean spaces for $n\geq 2$, known as vector interference alignment, has been studied by several researchers, e.g.,  \cite{maddah2008communication,jafar2008dfr,cadambe2007interference,cadambe2008degrees}.
In this method, at each receiver, a subspace is dedicated to
interference; then the signaling is designed such that all the
interfering signals are squeezed into the interference subspace.
Using this method, Cadambe and Jafar \cite{cadambe2007interference} showed that a $K$-user Gaussian Interference Channel (GIC) with varying channel gains could achieve the total DoF of
$\frac{K}{2}$. Since the assumption of varying channel gains is
unrealistic, particularly that all the gains should be known at the transmitters,  the practical application of these important theoretical results is limited.

Motahari et al. \cite{motahari2009real} settled the problem for the general scenario by proposing a new type of
interference alignment that can achieve $\frac{K}{2}$ DoF for almost all $K$-user real GIC with constant coefficients. This result was obtained by introducing a new type of interference alignment known as
real interference alignment. In this technique, tools from the field of Diophantine
approximation in number theory play a crucial role, see---Appendix. Studies such as \cite{cadambe2007interference, motahari2009real} showed that for a $K$-user $M$ antenna MIMO interference channel, the total DoF is equal to $\frac{KM}{2}$, whether the channel is constant or time varying/frequency selective.

In \cite{motahari2009real}, a scheme  similar  to ~\cite{bresler2010approximate} is used where both signal and interference are
received in a single communication dimension, but
unlike~\cite{bresler2010approximate}, the signal and interference are
not separated based on the received power level. \cite{motahari2009real} shows that
the properties of real numbers can be exploited to align signals and
achieve the full DoF of time invariant interference channels.

 Although \cite{motahari2009real} shows that the total
DoF of $\frac{4}{3}$ for the single antenna $2\times 2$, X channel can be achieved,
the MIMO X channel cannot be treated similarly. The MIMO X channel behaves differently compared with the $K$-user MIMO GIC. Although in the latter the total DoF is fully characterized for the case of equal number of antennas at all nodes, the corresponding problem in the former setup is still open. It is observed that  neither {``\textit {vector interference alignment"}}
nor {``\textit {real interference alignment"}} techniques can provide the necessary means to settle the problem individually. The aim of this paper is to introduce a new type of interference alignment, called \emph{layered interference alignment}, in which a similar approach to real interference alignment is used in conjunction with signal linear pre-coding (similar to vector alignment) to obtain optimal (in terms of DOF) signaling for the MIMO, $K\times 2$ and $2\times K$, X channels. Derivations rely on a new number theoretic measure estimates that are proved in this paper.

%%%%%%%%%%%%%% System Model%%%%%%%%%%%%%%%%%%%%%

\section{System Model}
\subsection{Notation}
Throughout this article, boldface upper-case letters, e.g., $\bf{H}$, are used to represent matrices. Matrix elements will be shown in brackets, e.g., $\bf{H}$ $= [h_{i,j}]$ for a set of values $i,j$. Vectors are shown using boldface italic lower-case letters, e.g., \mbox{\boldmath $v$}. Vector elements are shown inside parentheses, e.g., \mbox{\boldmath $v$} $= (v_1,v_2,...,v_{i})$. The transpose and conjugate transpose of a matrix $\bf{A}$ will be represented as $\bf{A}$ $\!\!^t$ and $\bf{A} ^{\dagger}$, respectively.  In general, the transmitted signal from the $k$th antenna of transmitter $i$, desired to be decoded at receiver $j$, is represented by $x_k^{i,j}$. At each antenna of transmitters in the X channel, a linear combination of all desired messages for different receivers will be transmitted. To simplify the derivations, with some misuse of notation, we define $x_k^{i}=\sum_{j}\beta_{j}x_k^{i,j}$, where $\beta_{j}$ is the weight of message $x_k^{i,j}$  for linear encoding at transmitter $i$. The transmitted vector signal at transmitter $i$ will be represented as  \mbox{\boldmath $x$}$^{i} = (x^i_1,x^i_2,...,x_k^{i})^t$. We use single superscript labelling for the indices of both transmitters and receivers, for example, \mbox{\boldmath $z$}$^i$ represents the noise vector at the receiver $i$. Single subscripts are used for the antenna labelling unless otherwise stated; for example, $y^{i}_{j}$ represents the received signal at the $j$th antenna of receiver $i$. The superscript pair $i,j$ represents the variable from transmitter $i$ to receiver $j$, and similarly the subscript pair $l,n$ represents the variable from antenna $l$ to antenna $n$. For example, $h^{i,j}_{l,n}$ represents the channel gain between the $l$th antenna of transmitter $i$ and the $n$th antenna of the receiver $j$. We use upper-case calligraphic alphabets to represent the set of constellation points such as  $\mathcal{U}$. The $M$ dimensional ring of integers is represented by $\mathbb{Z}^M$.

\subsection{$K$-Transmitter, 2-Receiver, $M$ Antenna X Channel}
A constant fully connected $K$-transmitter, 2-receiver MIMO Gaussian X channel is considered. This channel models a communication network with $K$ transmitters and two receivers. Each transmitter is equipped with $M$ antennas and wishes to communicate with both receivers, transmitting a dedicated message to each of them. Each of the receivers is also equipped with $M$ antennas. All transmitters share a common bandwidth. The channel outputs at the receivers are characterized by the following input-output relationship:
\begin{eqnarray*}
\mbox{\boldmath $y$}^{i}={\bf H}^{1,i}\mbox{\boldmath $x$}^{1}+{\bf H}^{2,i}\mbox{\boldmath $x$}^{2}+...+{\bf H}^{K,i} \mbox{\boldmath $x$}^{K}+\mbox{\boldmath $z$}^{i}
\end{eqnarray*}
where $i \in \{1,2\}$ is the receiver index, $k \in \{1,2,...,K\}$ is the transmitter index, \mbox{\boldmath $y$}$^{i}=(y^{i}_{1},y^{i}_{2},...,y^{i}_{M})^t$ is the $M \times 1$ output vector signal of the $i$th receiver, \mbox{\boldmath $x$}$^{j}=(x^{j}_{1},x^{j}_{2},...,x^{j}_{M})^{t}$ is the $M \times 1$ input vector signal of the $j$th transmitter, ${\bf H}^{j,i}=[h^{j,i}_{l,n}]$ is the $M \times M$ channel matrix between transmitter $j$ and receiver $i$, where $h_{l,n}^{j,i}$ specifies the channel gain from the $l$th antenna of the $j$th transmitter to the $n$th antenna of the $i$th receiver, and \mbox{\boldmath $z$}$^{i}=(z^{i}_1,z^{i}_2,...,z^{i}_M)^t$ is $M \times 1$ Additive White Gaussian Noise (AWGN) vector at receiver $i$. All noise terms are assumed to be independent and identically distributed (i.i.d.), zero mean, unit variance Gaussian random variables. It is assumed that each transmitter is subject to an average power constraint $P$, i.e.,
\begin{eqnarray*}
\mathbb{E}[(\mbox{\boldmath $x$}^{j})^{\dagger}(\mbox{\boldmath $x$}^{j})] \leq P
\end{eqnarray*}
where $\mathbb{E}[.]$ represents the expectation. As mentioned earlier, the transmitted signal from the $k$th antenna of transmitter $i$ desired to be decoded at receiver $j$ is represented by $x_k^{i,j}$. At each antenna of each transmitter, a linear combination of all desired messages for different receivers will be transmitted. Recall that $x_k^{i}=\sum_{j}\beta_{j}x_k^{i,j}$, where $\beta_{j}$ is the weight of message $x_k^{i,j}$ in the linear combination.

Let $P_{e}^{j,i}$ denote the probability of error for a message sent by transmitter $j$ to receiver $i$, i.e.,
\begin{eqnarray*}
 P_{e}^{j,i}=Pr\{W^{j,i}\neq \hat{W}^{j,i}\}
\end{eqnarray*}
where $W^{j,i}$ is the message sent by transmitter $j$ to receiver $i$ with the rate $R^{j,i}$ and $\hat{W}^{j,i}$ is the corresponding decoded message.

For a given power constraint $P$, a rate region $R(P)$ is determined by $R^{j,i}$'s. The closure of the set of all achievable rate tuples is called the capacity region of the channel with power constraint $P$ and is denoted by $\mathcal{C}(P)$. The notion of DoF is defined next.
\begin{definition}
To an achievable rate tuple $R(P) \in \mathcal{C}(P)$, one can correspond an achievable DoF of $d^{j,i}$ provided that
\begin{eqnarray*}
R^{j,i}= \frac{1}{2}d^{j,i}\log_{2}(P)+ \texttt{o}(\log_{2}(P)).
\end{eqnarray*}
 The set of all achievable DoF tuples is called the DoF region and is denoted by $\mathscr{D}$.
\end{definition}
\begin{definition}
The maximum sum rate or sum capacity of the $K$-transmitter, 2-receiver MIMO X channel is defined as
\begin{eqnarray*}
\textbf{C}_{\sum}(P)= \max_{R^{j,i} \in \mathcal{C}(P)}\sum_{i=1}^{2}\sum_{j=1}^{K} R^{j,i}.
\end{eqnarray*}
The maximum achievable sum DoF (or simply total DoF) is defined as
\begin{eqnarray*}
D= \max_{d^{j,i}\in \mathcal{D}}\sum_{i=1}^{2} \sum_{j=1}^{K} d^{j,i}.
\end{eqnarray*}
\end{definition}
In sequel, the notation $(K \times 2 ,M)$ X channel refers to $K$-transmitter, 2-receiver MIMO X channel with $M$ antennas at each transmitter/receiver.
\begin{figure}
\centering \scalebox{.8} {
\begin{pspicture}(0,0)(10,13)
%\psgrid

\psframe(7.1,3.35)(8.1,6.35)
\rput(8.2,4.5){\antenna}
\rput(8.2,6.5){\antenna}
\rput(7.6,5){$\vdots$}

\psframe(7.1,7)(8.1,10)
\rput(8.2,8.15){\antenna}
\rput(8.2,10.15){\antenna}
\rput(7.6,8.65){$\vdots$}

\psframe(3,0)(4,3)
\rput(4.1,1.15){\antenna}
\rput(4.1,3.15){\antenna}
\rput(3.5,1.65){$\vdots$}

\rput(3.5,4.5){$\vdots$}

\psframe(3,6)(4,9)
\rput(4.1,7.15){\antenna}
\rput(4.1,9.15){\antenna}
\rput(3.5,7.65){$\vdots$}

\psframe(3,9.5)(4,12.5)
\rput(4.1,10.65){\antenna}
\rput(4.1,12.65){\antenna}
\rput(3.5,11.15){$\vdots$}

\psline{->}(1.7,11.2)(2.5,11.2) \rput(2,11.4){\scriptsize \mbox{\boldmath $x$}$^{1}$}
\psline{->}(1.7,7.7)(2.5,7.7) \rput(2,7.9){\scriptsize \mbox{\boldmath $x$}$^{2}$}
\psline{->}(1.7,1.7)(2.5,1.7) \rput(2,1.9){\scriptsize \mbox{\boldmath $x$}$^{K}$}

\psline{->}(8.2,8.7)(9,8.7) \rput(8.4,8.9){\scriptsize \mbox{\boldmath $y$}$^{1}$}
\psline{->}(8.2,5.1)(9,5.1) \rput(8.4,5.3){\scriptsize \mbox{\boldmath $y$}$^{1}$}

\cnode[linecolor=white](4,11.2){.1}{T2}
\cnode[linecolor=white](4,7.5){.1}{T3}
\cnode[linecolor=white](4,1.7){.1}{T4}
\cnode[linecolor=white](7,8.9){.1}{R1}
\cnode[linecolor=white](7,4.2){.1}{R2}

\ncline[linestyle=solid]{->}{T2}{R1} \ncput*{\scriptsize  ${\bf{H}} ^{1,1}$}

\ncline[linestyle=dashed]{->}{T2}{R2} \ncput*{\scriptsize ${\bf{H}} ^{1,2}$}

\ncline[linestyle=solid]{->}{T3}{R1} \ncput*{\scriptsize  ${\bf{H}} ^{2,1}$}

\ncline[linestyle=dashed]{->}{T3}{R2} \ncput*{\scriptsize ${\bf{H}} ^{2,2}$}

\ncline[linestyle=solid]{->}{T4}{R1} \ncput*{\scriptsize  ${\bf{H}} ^{K,1}$}

\ncline[linestyle=dashed]{->}{T4}{R2} \ncput*{\scriptsize ${\bf{H}} ^{K,2}$}

\end{pspicture}
}
\caption{$K\times 2$, $M$ antenna X channel}\label{fig kx2}
\end{figure}
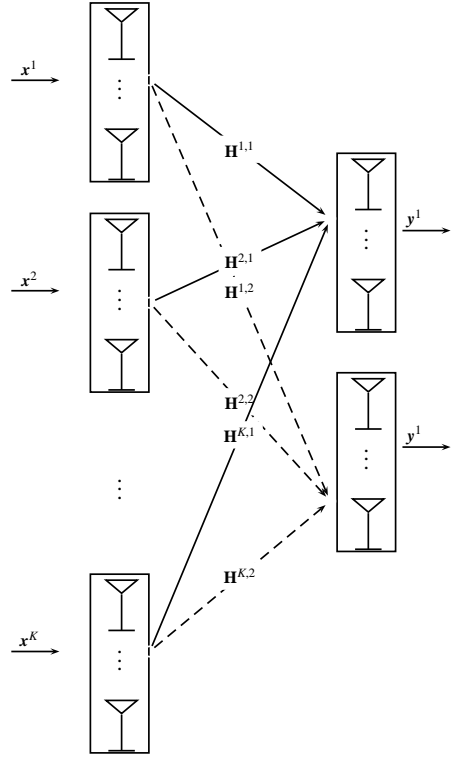

\subsection{2-Transmitter, $K$-Receiver, $M$ Antenna X Channel}
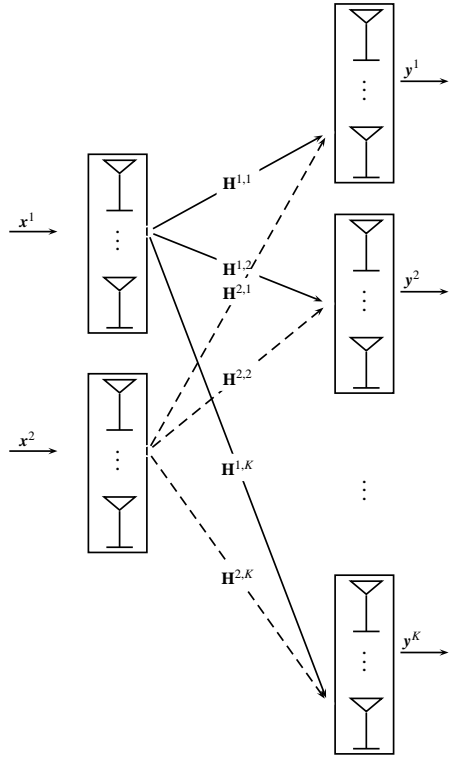
\begin{figure}
\centering \scalebox{.8} {
\begin{pspicture}(0,0)(10,13)
%\psgrid
\psframe(3,3.35)(4,6.35)
\rput(4.1,4.55){\antenna}
\rput(4.1,6.5){\antenna}
\rput(3.5,5){$\vdots$}

\psframe(3,7)(4,10)
\rput(4.1,8.20){\antenna}
\rput(4.1,10.15){\antenna}
\rput(3.5,8.65){$\vdots$}

\psframe(7.1,0)(8.1,3)
\rput(8.2,1.20){\antenna}
\rput(8.2,3.15){\antenna}
\rput(7.6,1.65){$\vdots$}

\rput(7.6,4.5){$\vdots$}

\psframe(7.1,6)(8.1,9)
\rput(8.2,7.20){\antenna}
\rput(8.2,9.15){\antenna}
\rput(7.6,7.65){$\vdots$}

\psframe(7.1,9.5)(8.1,12.5)
\rput(8.2,10.70){\antenna}
\rput(8.2,12.65){\antenna}
\rput(7.6,11.15){$\vdots$}

\psline{->}(1.7,8.7)(2.5,8.7) \rput(2,8.9){\scriptsize \mbox{\boldmath $x$}$^{1}$}
\psline{->}(1.7,5.05)(2.5,5.05) \rput(2,5.25){\scriptsize \mbox{\boldmath $x$}$^{2}$}

\psline{->}(8.2,1.7)(9,1.7) \rput(8.4,1.9){\scriptsize \mbox{\boldmath $y$}$^{K}$}
\psline{->}(8.2,7.7)(9,7.7) \rput(8.4,7.9){\scriptsize \mbox{\boldmath $y$}$^{2}$}
\psline{->}(8.2,11.2)(9,11.2) \rput(8.4,11.4){\scriptsize \mbox{\boldmath $y$}$^{1}$}

\cnode[linecolor=white](4,8.7){.1}{T1}
\cnode[linecolor=white](4,5.05){.1}{T2}
\cnode[linecolor=white](7,10.35){.1}{R1}
\cnode[linecolor=white](7,7.5){.1}{R2}
\cnode[linecolor=white](7,0.85){.1}{Rk}

\ncline[linestyle=solid]{->}{T1}{R1} \ncput*{\scriptsize ${\bf{H}} ^{1,1}$}
\ncline[linestyle=solid]{->}{T1}{R2} \ncput*{\scriptsize ${\bf{H}} ^{1,2}$}
\ncline[linestyle=solid]{->}{T1}{Rk} \ncput*{\scriptsize ${\bf{H}} ^{1,K}$}

\ncline[linestyle=dashed]{->}{T2}{R1} \ncput*{\scriptsize ${\bf{H}} ^{2,1}$}
\ncline[linestyle=dashed]{->}{T2}{R2} \ncput*{\scriptsize ${\bf{H}} ^{2,2}$}
\ncline[linestyle=dashed]{->}{T2}{Rk} \ncput*{\scriptsize ${\bf{H}} ^{2,K}$}
\end{pspicture}
}
\caption{$2\times K$, $M$ antenna X channel}\label{fig 2xk}
\end{figure}

A fully connected 2-transmitter, $K$-receiver MIMO Gaussian X channel is considered. Transmitters and receivers are equipped with $M$ antennas (see Figure \ref{fig 2xk}). The channel outputs at the receivers are characterized by the following input-output relationships:
\begin{center}
\mbox{\boldmath $y$}$^{i}={\bf H}^{1,i}$ \mbox{\boldmath $x$}$^{1}+{\bf H}^{2,i}$\mbox{\boldmath $x$}$^{2}+$\mbox{\boldmath $z$}$^{i}$
\end{center}
where $i \in \{1,2,...,K\}$ is the receiver index and \mbox{\boldmath $z$}$^{i}=(z^{i}_1,z^{i}_2,...,z^{i}_M)^t$ is  the $M \times 1$ AWGN vector at receiver $i$. Similar to the $K\times 2$, MIMO X channel, sum capacity and DoF region for $2 \times K$, MIMO X channels can be defined. In the sequel, the notation $(2 \times K ,M)$ X channel refers to constant channel gain, $2$-transmitter, $K$-receiver MIMO X channel with $M$ antennas at each transmitter/receiver.

\section{Main Contributions and Discussion}\label{sec main theorem}
\subsection{Main Contributions}
In this article, the total DoF of the following channels are characterized: \\
1. ($2 \times K, M$) X channel with constant real or complex channel realization. \\
2. ($K \times 2, M$) X channel with constant real or complex channel realization.\\
It is observed that the duality/reciprocity holds for the DoF of this class of X channels, i.e., if the role of transmitters is  interchanged with that of receivers, the total DoF will be conserved. The technique used in this article, named layered interference alignment,  benefits from a linear pre-coding similar to vector interference alignment at transmitters, in conjunction with a number theoretic technique similar to that of real alignment using rational dimensions at transmitters. A new mathematical tool is introduced to empower the use of joint processing and mutual decoding among the receiver antennas to achieve the total fractional DoF of each desired message. The main results can be stated as follows:

\begin{theorem}\label{main 1}
The total DoF of ($K\times 2$, $M$) X channel is $\frac{2KM}{K+1}$ for
almost all channel realizations.
\end{theorem}
\begin{theorem}\label{main 2}
The total DoF of the ($2\times K$, $M$) X channel is  $\frac{2KM}{K+1}$ for
almost all channel realizations.
\end{theorem}
This implies that when the base for comparison is the DoF,  ($2\times K$, $M$) and ($K\times 2$, $M$) X channels are dual/reciprocal.

\begin{theorem}\label{main 3}
The total DoF of the ($2\times K$, $M$) X channel and its dual, the ($K\times 2$, $M$) X channel with complex and time invariant channel coefficients, is  $\frac{4KM}{K+1}$ for
almost all channel realizations.
\end{theorem}
 This is twice that of the same channel with real channel gains. Note that the DoF for complex channel realizations should be defined as  half  of its value for real channels, since the complex case uses two dimensions for each transmission. This implies that the total DoF per transmit dimension is the same as real channel realization, which is equal to $\frac{2KM}{K+1}$.

A crucial ingredient in proving these theorems is the connection with the `size' estimates of sets of real or complex numbers having certain approximation properties. Such approximation properties are modelled in the linear forms setup.  The Khintchine--Groshev type theorems play a central role in determining the `size' of such sets by means of convergence or divergence of certain series which entirely depend upon the approximation error of the linear forms. For such linear forms  we establish Khintchine--Groshev type theorems in the Appendix.

 Before getting into details of layered interference alignment, we need to review some basics of transmit signal design using rational dimensions and a simple decoder design for the real interference alignment. We will go through some basic examples that show how the conventional interference alignment techniques fall short in some simple channels. We will go through the deployment of the layered interference alignment for $K \times 2$ and $2 \times K$ X channels.

\subsection{Interference Alignment}
In the following, we will discuss the general encoder and decoder design for aligning interference in X channels. A single layer constellation is used to modulate data streams at each transmitter. Despite its simplicity, it is powerful enough to support interference alignment, and achieve the DoF of the X channel. Prior to deriving the main results, the performance of a typical decoding technique is analysed. Throughout this paper, we will rely on these results, in conjunction with a special form of Khintchine-Groshev type theorem. It is noteworthy that in \cite{motahari2009real}, the authors showed for constant real channel gains, the total DoF of $\frac{4}{3}$ is achievable for a $2 \times 2$, SISO X channel.

\subsection{Transmission using Rational Dimensions}
To simplify notations, the desired message for the first receiver is noted as \mbox{\boldmath $u$}$^j$=$(u^j_1,u^j_2,...,u^j_M)^t$, and the desired message for the second receiver is noted as \mbox{\boldmath $v$}$^j$ $=(v^j_1,v^j_2,...,v^j_M)^t$.

Transmitter $j$ selects two constellations, $\mathcal{U}^j$ and  $\mathcal{V}^j$, to
send data stream $j$ to both receivers. The corresponding constellation points are chosen from the set of integers, i.e., $\mathcal{U}^j\subset
\mathbb{Z}^M$ and $\mathcal{V}^j\subset
\mathbb{Z}^M$. It is assumed that $\mathcal{U}^j$ and $\mathcal{V}^j$ are bounded sets. Hence, there is a constant $Q$ such
that $\mathcal{U}^j\subset [-Q,Q]$ and $\mathcal{V}^j\subset [-Q,Q]$ intervals. The maximum  cardinality of $\mathcal{U}^j$ and $\mathcal{V}^j$, which limits the rate of data stream $j$,
is denoted by $|\mathcal{X}^j|=\max\{|\mathcal{U}^j| , |\mathcal{V}^j|\}$.
This design corresponds to the case where all integers between $-Q$ and $Q$ are selected, which, in spite of its simplicity, is capable of achieving the total DoF for several channels.

Having formed the constellation, the transmitter constructs two random codebooks for data stream $j$ with rates $R^{j,1}$ and $R^{j,2}$ to be received by the first and the second receivers, respectively. This can be accomplished by choosing a probability distribution on the input alphabets. A uniform distribution is used  for the sake of simplicity. Note that, since the input constellation is
symmetrical by assumption, the expectation of the uniform distribution is zero. The power consumed by the data stream $j$ can be bounded as $Q^2$. Even though this bound is not tight, it does not decrease the performance of the system as far as the DoF is concerned. The transmit signal at the $l$th antenna of transmitter $j$ can be represented as
\begin{equation*}
x^j_l=a_{l}^{j}u^j_l+b_{l}^{j}v^j_l.
\end{equation*}
where $u^j_l$ contains the partial information in data stream $j$ that is intended to the first receiver and is being transmitted by the $l$th antenna of transmitter $j$. Accordingly, $v^j_l$ presents the part of the information for data stream $j$ that is desired at the second receiver and is being transmitted by the $l$th antenna of transmitter $j$.

Real numbers $a_{l}^{j}$ and $b_{l}^{j}$ are rationally independent, i.e., the equation $a_{l}^{j}x_1+b_{l}^{j}x_2=0$ has no rational solutions for each $j \in \{1,2,...,K\}$ and $l \in \{1,2,...,M\}$. This independence is because a unique map from constellation points to the
message sets is required. Reliance on this independence means that any real number $ x^j_l$ belonging to the set of constellation points is uniquely decomposable as $ x^j_l=a_{l}^{j}u^j_l+b_{l}^{j}v^j_l$. Observe that if there is another possible
decomposition $x^j_l=\hat{a}_{l}^{j}u^j_l+\hat{b}_{l}^{j}v^j_l$, then it forces $\hat{a}_{l}^{j}$ and $\hat{b}_{l}^{j}$ to be rationally dependent.

With the above method, each transmitter forms its transmit data stream \mbox{\boldmath $\hat{x}$}$^j$ $=(a_{l}^{j}u^{j}_l+b_{l}^{j}v^j_l)$ for $l=1,2,...,M$. To adjust the power, the transmit signal is multiplied by a constant $A$, i.e., the transmit signal is \mbox{\boldmath $x$}$^j=A$ \mbox{\boldmath $\hat{x}$}$^j$.

\subsection{Recovering the Mixed Signal in Rational Dimensions}
After rearrangement of the interfering term, the received signal can be represented as
\begin{equation}\label{received signal}
 y=\hat{g}^0u^0+\hat{g}^1I^1+\ldots+\hat{g}^mI^m+z.
\end{equation}
Hereafter, we consider $\hat{g}^0=g^0$ to unify the notation. Next, the decoding scheme used to decode $u^0$ from $y$   is explained. It is worth noting that if the receiver is interested in more than one data stream, then it performs the
same decoding procedure for each data stream.

At the receiver, the received signal is first passed through a hard decoder.
The hard decoder maps the received point
$\hat{\mathcal{U}}=g^0\mathcal{U}^0+\hat{g}^1\mathcal{I}^1+\ldots+\hat{g}^m\mathcal{I}^m$ to the nearest point in the constellation. This changes the continuous channel to a discrete-input, discrete-output channel in which the
input symbols are from the transmit constellation $\mathcal{U}^0$ and the output symbols are from the received
constellation.

Note that $\mathcal{I}^j$ is the constellation due to single or multiple data streams. Since it is assumed that in the latter case there is a linear combination of multiple data streams with integer coefficients, it can be concluded that
$\mathcal{I}^j\subset \mathbb{Z}$ for $j\in\{1,2,\ldots,m\}$.

To bound the performance of the decoder, it is assumed that the received constellation has the property that there
is a many-to-one map from $\hat{\mathcal{U}}$ to $\mathcal{U}^0$. This in fact implies that if there is no additive
noise in the channel, then the receiver can decode the data stream with zero error probability. This property is
called property $\Gamma$. It is assumed that this property holds for all received constellations. To satisfy this
requirement at all receivers, usually a careful transmit constellation design is needed at all transmitters, which will be explained next.

Let $d_{\text{min}}$ denote the minimum distance in the received constellation. Having property $\Gamma$, the
receiver passes the output of the hard decoder through the many-to-one mapping from $\hat{\mathcal{U}}$ to $\mathcal{U}^0$.
The output is called $\hat{u}^0$. Now, a joint-typical decoder can be used to decode the data stream from a block of
$\hat{u}^0$. To calculate the achievable rate, the error probability, i.e., $P_e=Pr\{\hat{U}^0\neq U^0\}$, is bounded as
\begin{eqnarray}\label{error probability}
P_e &\leq Q\left(\frac{d_{\text{min}}}{2\sigma}\right)\leq
\exp\left({-\frac{ d_{\text{min}}^2}{8\sigma^2}}\right).
\end{eqnarray}

\begin{definition}[Noise Removal]\label{NR}
A receiver can completely remove the noise if the minimum distance
between the received constellation points is greater than $\sqrt{N}$,
where $N$ is the noise variance \cite{motahari2009real}.
\end{definition}

Now $P_e$ can be used to lower bound the achievable rate. Etkin and Ordentlich \cite{Etkin-Ordentlich} used Fano's
inequality to obtain a lower bound on the achievable rate, which is tight in high Signal-to-Noise Ratio (SNR) regimes. Following similar
steps, one obtains
\begin{eqnarray}
R   && =I(\hat{u}^0,u^0)\nonumber\\
    && =H(u^0)-H(u^0|\hat{u}^0)\nonumber\\
    && \stackrel{a}{\geq} H(u^0)-1-P_e\log |\mathcal{U}^0|\nonumber\\
    && \stackrel{b}{\geq} \log |\mathcal{U}^0|-1-P_e\log |\mathcal{U}^0| \label{lower bound on R}
\end{eqnarray}
where (a) follows from Fano's inequality and (b) follows from the fact that $u^0$ has  uniform distribution. To
have a multiplexing gain of at least $d$, $|\mathcal{U}^0|$ needs to scale as $\text{SNR}^{d}$. Moreover, if $P_e$ scales as
$\exp\left(\text{SNR}^{-\epsilon}\right)$ for an $\epsilon>0$, then it can be shown that $\frac{R}{\log \text{SNR}}$ approaches $d$ at high SNR regimes.

\subsection{Main Ideas and Basic Examples}

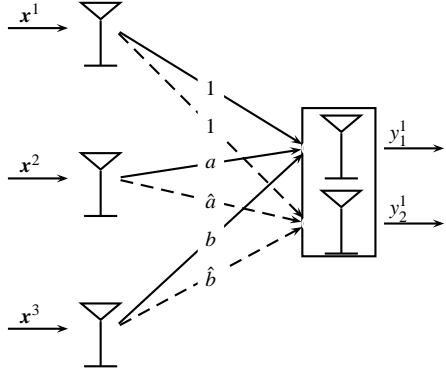
\begin{figure}
\centering \scalebox{1} {
\begin{pspicture}(0,0)(7,6)
\psgrid[gridcolor=white,subgridcolor=white,gridlabels=0]
\rput(3,1){\antenna} \rput(3,3){\antenna} \rput(3,5){\antenna}
\psframe(5.1,1.35)(6.1,3.35) \rput(6.2,2.5){\antenna}
\rput(6.2,3.5){\antenna}

\psline{->}(1.2,4.4)(2,4.4) \rput(1.5,4.6){\scriptsize \mbox{\boldmath $x$}$^{1}$}
\psline{->}(1.2,2.4)(2,2.4) \rput(1.5,2.6){\scriptsize \mbox{\boldmath $x$}$^{2}$}
\psline{->}(1.2,0.4)(2,0.4) \rput(1.5,0.6){\scriptsize \mbox{\boldmath $x$}$^{3}$}

\psline{->}(6.2,1.8)(7,1.8) \rput(6.4,2){\scriptsize $y_2^1$}
\psline{->}(6.2,2.8)(7,2.8) \rput(6.4,3){\scriptsize $y_1^1$}

\cnode[linecolor=white](2.6,4.4){.1}{T2}
\cnode[linecolor=white](2.6,2.4){.1}{T3}
\cnode[linecolor=white](2.6,0.4){.1}{T4}
\cnode[linecolor=white](5.2,2.8){.1}{R1}
\cnode[linecolor=white](5.2,1.8){.1}{R2}

\ncline[linestyle=solid]{->}{T2}{R1} \ncput*{\scriptsize $1$}

\ncline[linestyle=dashed]{->}{T2}{R2} \ncput*{\scriptsize $1$}

\ncline[linestyle=solid]{->}{T3}{R1} \ncput*{\scriptsize $a$}

\ncline[linestyle=dashed]{->}{T3}{R2} \ncput*{\scriptsize $\hat{a}$}

\ncline[linestyle=solid]{->}{T4}{R1} \ncput*{\scriptsize $b$}

\ncline[linestyle=dashed]{->}{T4}{R2} \ncput*{\scriptsize $\hat{b}$}

\end{pspicture}
} \caption{SIMO multiple access channel}\label{fig Mac}
\end{figure}

\subsubsection*{Single-Input Multiple-Output (SIMO) Multiple Access Channel}
A Multiple Access Channel (MAC) with three single antenna users and a
2-antenna receiver is shown in Figure \ref{fig Mac}. The channel can
be modelled as
\begin{equation}
\left\{\begin{array}{rl}
y_1^1\!\!\!\!\! &=x^{1}+ax^{2}+bx^{3}+z_1\\
y_2^1\!\!\!\!\!&=x^{1}+\hat{a}x^{2}+\hat{b}x^{3}+z_2
\end{array}  \right.
\end{equation}
where all channel gains are constant, real numbers.

Since the capacity region of this channel is fully characterized, it can be easily shown
that the total DoF is 2. Vector interference alignment falls short of achieving this DoF, as transmitters are equipped with a single antenna. The
naive application of real interference alignment results in a similar shortcoming. To see this, let us assume that
all three users communicate with the receiver using a single data
stream. The data streams are modulated by the constellation
$\mathcal{U}=A(-Q,Q)_{\mathbb{Z}}=\{\text{all integers between $-Q$ and $Q$}\}$, where $A$ is a factor controlling the minimum distance of the received constellation.

The received constellation, which is a set of points in a two-dimensional space, consists of points $(v,\hat{v})$ such that $v=A(u^1+au^2+bu^3)$
and $\hat{v}=A(u^1+\hat{a}u^2+\hat{b}u^3)$, where $u^i$ 's are members of $\mathcal{U}$. Let us choose two sets of distinct points $(v_1,\hat{v_1})$ and $(v_2,\hat{v_2})$ in the
received constellation. The Khintchine-Groshev theorem provides
a lower bound on any linear combination of integers. It also
provides some bound on the distance between any integer vector and the linear combination of
 rationally independent vectors. Using the Khintchine--Groshev theorem (Theorem \ref{KG}  in \S A of appendix)for $m=2, n=1$, one can obtain $d_{\min}\approx
\frac{A}{Q^2}$, where $d_{\min}$ is the minimum distance in the
received constellation, for precise calculation of min distance we refer to \cite[\S A]{motahari2009real}. %This theorem will be discussed and proved in details in an appendix.

By using the noise removal definition (Def. \ref{NR}) and assuming unit variance for the Gaussian noise, the noise can be removed if $d_{\min}=1$. Hence, it is sufficient to have $A\approx Q^2$. In a noise-free
environment, each receiver antenna can decode the three messages if
there is a one-to-one mapping from the received constellation to the transmit
constellations. Mathematically, one can satisfy the separability condition by enforcing the following:
Each received antenna is able to decode all three messages if the channel coefficients associated with that antenna are rationally independent.
In the above multiple access channel, for instance, the receiver can decode all messages by using the signal from the first antenna if
$u^1+au^2+bu^3=0$ has no non-trivial solution in integers for $u^1$, $u^2$ and $u^3$.

User $i$'s rate is equal to $R^i=\log(2Q-1)$. Because of the power constraint,
$P=A^2Q^2$. It was shown earlier that $A\approx Q^2$. Therefore, $P\approx
Q^6$. Hence,
\begin{equation}
d^i=\lim_{P\rightarrow\infty}\frac{R^i}{0.5\log P}=\frac{1}{3}.
\end{equation}
If all three messages are decoded, the achievable DoF for this channel would  be 1, while the total DoF is proved to be 2. In \cite{motahari2009real}, authors deployed the real interference alignment technique to
achieve the total DoF for the SISO multiple access channel, but this scheme falls short for the general MIMO channel.

Motivated by above shortcomings, a new alignment scheme called layered interference alignment is proposed to achieve the total DoF of this channel and a class of MIMO channels. This technique, in general, combines  vector and  real interference alignment techniques in a subtle way to enjoy the benefits of multiple antennas at both transmit and receive sides. The  SIMO multiple access channel considered in this section has no room for vector alignment. Above example helps to understand the difference between the real and the layered interference alignment. Concretely, the  above shortcomings can be resolved using joint decoding of the received signals by incorporating a new Khintchine-Groshev type theorem. This theorem bounds the $d_{\min}$ based on the size of the input constellation and the number of antennas. These results are backed up by Theorem \ref{thm1}, which will be discussed in detail in section \ref{new-problem}. To use these mathematical results, one must provide an algorithm at receivers for simultaneous decoding.

\subsubsection{Joint Processing of Received Data Streams}
This operation is composed of the followings:

1. Each receiver first normalizes its received data streams in order to have the unity coefficient for a specified favourite message at all receiver antennas.

2. After normalization, each receiver uses the results of Theorem \ref{thm1} to simultaneously decode each message from all received streams at each of the $M$ antennas.

The same procedure will be reapplied for all other desired messages.

In the multiple access channel example, joint
decoding is employed at both receiver antennas. User $i$'s rate is
$R^i=\log(2Q-1)$. Because of the power constraint, we have $P=A^2Q^2$.
Applying Theorem \ref{thm1} (for $m=n=2 $) and satisfying the noise removal
assumption results in $A\approx Q^{0.5}$. Therefore, $P\approx
Q^3$. So
\begin{equation}
d^i=\lim_{P\rightarrow\infty}\frac{R^i}{0.5\log P}=\frac{2}{3}.
\end{equation}
Using the above method to decode each of the three messages, each of which has the DoF of $\frac{2}{3}$, results in the total
DoF of 2, which is the desired result. In the rest of this article, we incorporate layered interference alignment in its full potential, i.e., having the vector and the real interference alignment together with joint processing, to achieve the total DoF for ($K\times 2,M$) and ($2\times K,M$) X channels.

\subsubsection{Complex Coefficients}
Unlike the MAC, it can be easily seen that the total DoF of the X channel with complex coefficients cannot
be achieved by pairing \cite{maddah2010degrees}. In this case, using layered interference alignment
requires a new joint processing bound, which will be discussed separately in Section \ref{complex-case}. This new theorem leaves the encoding and decoding methods intact and provides the required tools to analyze the performance of the layered interference alignment for the constant complex channel gains. It will be observed that this extension to the layered interference alignment technique will achieve the total DoF of $\frac{4KM}{K+1}$  for both ($K\times 2,M)$ and ($2\times K,M)$ X channels with constant complex channel gains. This is twice the DoF of the same channels with real channel coefficients.

%It worths emphasizing that, similar to the real alignment \cite{motahari2009real}, all the results presented in this paper are based on the separability
%conditions and Khintchine-Groshev type theorems. Therefore, all
%results are valid for almost all channel realizations.

\section{DoF of ($K \times 2$, $M$) X Channel with Constant Real Channel Gains}
%In the MIMO setup, it is required to extend the Khintchine-Groshev theorem to linear forms over vector space. It may be expected that such extension can be solved under simultaneous Diophantine approximation. This, however,  cannot be used directly to characterize the total DOF of MIMO X channel. On the other hand, it tunrs out that using an approach similar to the original proof of the Khintchine-Groshev theorem for vectors can be used to tackle the problem. This new tool enables the use of joint processing/simultaneous decoding at all receive antennas. It observed that such joint decoding is  crucial in the approximation of sets of numbers in higher dimensions.

In this section, we describe the encoding and decoding procedures which can achieve the total DOF of ($K \times 2$, $M$) X channel.

\subsection{Encoding}
The $i$th transmitter sends two sets of messages,
\mbox{\boldmath $u$}$^i$ $=(u_{1}^i,u_{2}^i,...u_{M}^i)^t$ and \mbox{\boldmath $v$}$^i$ $=(v_{1}^i,v_{2}^i,...v_{M}^i)^t$.
It is preferable to decode these at receivers 1 and 2, respectively. The transmitter selects its modulation points from $\mathcal{U}=A(-Q,Q)_{\mathbb{Z}}$ and
$\mathcal{V}=A(-Q,Q)_{\mathbb{Z}}$ for $u_{l}^i$ and $v_{l}^i$, $l=1,2,..,M$, accordingly. $A$ is a constant factor
that controls the minimum distance of the received constellation.

The transmit directions are first chosen in such a way that the interfering signals at both receivers are aligned. To this end, two $M \times M$ matrices ${\bf{I}} ^{1}$ and ${\bf{I}} ^{2}$ are fixed at receivers 1 and 2, respectively. ${\bf{I}}^{1}$ and ${\bf{I}}^{2}$ can be used to design the transmit signals. For instance, the
$i$th transmitter uses the following signal for data transmission:
\begin{equation}
\mbox{\boldmath $x$}^i=({\bf{H}} ^{i,2})^{-1} {\bf{I}}^{2}\mbox{\boldmath $u$}^{i}+({\bf{H}} ^{i,1})^{-1} {\bf{I}} ^{1}{\mbox{\boldmath $v$} ^{i}}.
\end{equation}
\subsection{Decoding}
The corresponding receive signals are
\begin{equation}
\left\{\begin{array}{rl}
{\mbox{\boldmath $y$} ^{1}}\!\!\!\!\!&=\sum_{i=1}^{K} ({\bf{H}} ^{i,1})({\bf{H}} ^{i,2})^{-1} {\bf{I}} ^{2}{\mbox{\boldmath $u$}^ {i}}+{\bf{I}}^{1}\sum_{i=1}^{K} {\mbox{\boldmath $v$}^ {i}}+{\mbox{\boldmath $z$} ^{1}}\\
{\mbox{\boldmath $y$}^ {2}}\!\!\!\!\!&=\sum_{i=1}^{K} ({\bf{H}} ^{i,2})({\bf{H}} ^{i,1})^{-1} {\bf{I}} ^{1}{\mbox{\boldmath $v$}^ {i}}+{\bf{I}} ^{2}\sum_{i=1}^{K} {\mbox{\boldmath $u$} ^i}+{\mbox{\boldmath $z$} ^{2}},
\end{array}  \right.
\end{equation}
where \mbox{\boldmath $z$}$^{1}$ and \mbox{\boldmath $z$}$^{2}$ are independent Gaussian random vectors with identity covariance matrices.
At the $l$th antenna of the first receiver,
\begin{equation}
 y_{l}^1=\sum_{i=1,...,K}\sum_{j=1,...,M}{g _{l,j}^{i} u_{j}^i}+\sum_{j=1,...,M}{\eta_{lj}\varGamma_{j}}+z_{l}^1
\end{equation}
where $g _{l,j}^{i}$ is the receive gain (coefficient) for each $u_{j}^i$ observed at the $l$th antenna, and $\eta_{i,j}$ is the $i$th row, $j$th column component of matrix ${\bf I} ^1$ and $\varGamma_{j}$ is defined as
$\varGamma_{j}$=$\sum_{i=1}^{K}{v_{j}^i}$. Similarly, at the $l$th antenna of the second receiver, we have
\begin{equation}
 y_{l}^2=\sum_{i=1,...,K}\sum_{j=1,...,M}{\hat{g} _{l,j}^{i} v_{j}^i}+\sum_{j=1,...,M}{\lambda_{l,j}\Theta_{j}}+z_{l}^2,
\end{equation}
where ${\bf I} ^2$=$[\lambda_{i,j}]$ and $\Theta_{j}$=$\sum_{i=1}^{k}{u_{j}^i}$.

The first receiver can decode a message, say $u_{1}^1$, from the receive signals using the following algorithm. It first  normalizes the receive signal to set the coefficients of $u_{1}^1$ at all antennas to unity. Next, joint processing is applied to decode $u_{1}^1$. Theorem \ref{thm1} allows the minimum
distance to be approximated by $d_{\min}$=$AQ^{-k}$, see--Remark 1. Hence, setting $A\approx Q^k$ is sufficient to conclude $d_{\min}\approx 1$, which in turn results in noise removal from the received signal. Putting this together results in $P\approx Q^{2(k+1)}$. At the first receiver, one can obtain the following DoF for $u_{1}^1$:
\begin{equation}
 d^{1,1}=\lim_{P \rightarrow \infty}{ \frac{(R^{1,1}=\log{(2Q-1)})}{0.5\log{P}}}=\frac{1}{K+1}.
\end{equation}
This technique can be applied to all other partial messages at the first receiver. In the second receiver,  the same method will be applied for all
$v_{j}^i$, resulting in the same DoF for the second receiver. Finally, it is possible to decode $KM$ different messages at each receiver, which  results in the total DoF of $\frac{2KM}{K+1}$. This achieved DoF meets the upper bound mentioned in \cite{cadambe2008degrees}.

\section{DoF of ($2 \times K,M$) $X$ Channel with Constant Real Channel Gains}
In the following, we will show that the total DoF of ($2 \times K,M$) antenna X channel with constant real channel gains is the same as the DoF of ($K \times 2,M$) X channel, which is equal to $\frac{2KM}{K+1}$.

\subsection{Encoding}
The first transmitter sends the messages
\mbox{\boldmath $u$}$^j$ $=(u_{1}^j,u_{2}^j,...u_{M}^j)^t$ for $j=1,..,K$, and the second transmitter sends the messages \mbox{\boldmath $v$}$^j$ $=(v_{1}^j,v_{2}^j,...v_{M}^j)^t$; where
it is desired that \mbox{\boldmath $u$}$^{j}$ and \mbox{\boldmath $v$}$^{j}$ to be decoded at receiver $j$. The transmitter selects its modulation points from $\mathcal{U}=A(-Q,Q)_{\mathbb{Z}}$ and
$\mathcal{V}=A(-Q,Q)_{\mathbb{Z}}$ for $u_{l}^{j}$ and $v_{l}^j$, $l=1,2,...,M$, respectively, where $A$ is a constant factor that controls the minimum distance of the received constellation.

Similarr to the case of ($K \times 2$, $M$) X channel, the transmit directions are first chosen in such a way that the interfering signals at both receivers are aligned. To this end,  matrices ${\bf{I}}^{i}$, each of dimension $M\times M$, are fixed at receiver $i$, where ${\bf{I}}^{i}$'s is used to extract the transmit signals from all transmitters. The  goal at the $i$th receiver is
\begin{equation}
\mbox{\boldmath $y$}^{i}={\bf{H}} ^{1,i}\rho^{i}\mbox{\boldmath $u$}^{i}+{\bf{H}} ^{2,i}\zeta^{i} \mbox{\boldmath $v$} ^{i}+\sum_{{j=1} \& i\neq j}^{K} {\bf{I}} ^{i}+\mbox{\boldmath $z$} ^{i}.
\end{equation}
To obtain $\rho$ and $\zeta$,  the following solution is proposed:
\begin{displaymath}
    \left\{
     \begin{array}{lr}
      {\bf{H}} ^{1,i}\rho^{j}={\bf{H}} ^{2,i}\zeta^{j+1}  & j \not \in \{ i,i-1,K \}\\
      {\bf{H}} ^{1,i}\rho^{j}={\bf{H}} ^{2,i}\zeta^{j+2}  &  j= i-1\\
      {\bf{H}} ^{1,i}\rho^{j}={\bf{H}} ^{2,i}\zeta^{1}  &  j=K \ \& \  i \neq 1\\
      {\bf{H}} ^{1,i}\rho^{j}={\bf{H}} ^{2,i}\zeta^{2}  &  j=K \  \& \  i=1.
     \end{array}
   \right.
\end{displaymath}
 Using the above signal space design results in
 \begin{displaymath}
   {\bf{I}} ^{j}= \left\{
     \begin{array}{lr}
      {\bf{H}} ^{1,i}\rho^{j} ({\mbox{\boldmath $u$} ^{j}}+{\mbox{\boldmath $v$} ^{j+1}}) & j \not \in \{ i,i-1,K \}\\
      {\bf{H}} ^{1,i}\rho^{j} ({\mbox{\boldmath $u$} ^{j}}+{\mbox{\boldmath $v$} ^{j+2}})  &  j= i-1\\
      {\bf{H}} ^{1,i}\rho^{j} ({\mbox{\boldmath $u$} ^{j}}+{\mbox{\boldmath $v$} ^{1}})  &  j=K \ \&  \ i \neq 1\\
      {\bf{H}} ^{1,i}\rho^{j} ({\mbox{\boldmath $u$} ^{j}}+{\mbox{\boldmath $v$} ^{2}})  &  j=K \ \& \  i=1.
     \end{array}
   \right.
\end{displaymath}
\subsection{Decoding}
Using this signaling scheme, the received signal at the $l$th antenna of receiver $j$ can be expressed as
\begin{equation}
 y_{l}^j=\sum_{i=1,...,M}{\sigma _{l,i}u_{j}^i}+\sum_{i=1,...,M}{\lambda_{l,i}v_{j}^i}+\sum_{{i=1},i\neq j}^{K}\sum_{n=1}^{M}{I^{i}_{n}}+z_{l}^j,
\end{equation}
where $\sigma _{l,i}$ and $\lambda_{l,i}$ are constant coefficients representing the combined effects of all the channel gains for $u_{j}^i$ and $v_{j}^i$, respectively.

Now, applying the joint processing technique  at each antenna, results in receiving  the linear combination of  $2M$ desired partial messages ($M$ for \mbox{\boldmath $u$} and $M$ for \mbox{\boldmath $v$}) added to $M(K-1)$ interference terms. For any message $u_{j}^i$ at the $i$th antenna of receiver $j$, we use the joint processing among all the $M$ antennas. After normalizing, using Theorem \ref{thm1}, this results in
\begin{equation}
 d^{i,j}=\lim_{P \rightarrow \infty} \frac{\log{(2Q-1)}}{0.5\log{P}}=\frac{1}{K+1}
\end{equation}

The same argument is valid  for $v_{j}^i$, so it is concluded that the total DoF of $\frac{2KM}{K+1}$ is achievable.
It is observed that ($2 \times K,M$) and ($K \times 2,M$) X channels act reciprocal/dual in the sense of DOF. Here, for both ($2 \times K,M$) and ($K \times 2,M$) X channels  the achievability part is proved, since in \cite{cadambe2008degrees}, it is shown that the total DoF for both ($2 \times K,M$) and ($K \times 2,M$) X channels are upper bounded by $\frac{2KM}{K+1}$. Therefore, it can be concluded that the proposed schemes achieve the maximum DoF of these channels.

\section{Complex Coefficients Cases}
Let us consider the ($K \times 2, M$) X channel. It is shown in the previous section that the upper bound on the total DoF of $\frac{2KM}{K+1}$ is achievable for this channel when the channel gains are real. Needless to say, the result is also applicable to channels with complex coefficients.  The real and imaginary parts of the input and the output can be paired. This converts the channel to $2K$ virtual transmitters and 4 receivers. It can be seen that applying  Theorem \ref{thm1} does not achieve the upper bound on the DoF in this case.

To solve the issue, we will make an extension to  Theorem \ref{thm1} for complex channel realizations,  which can be used in characterizing the total DoF of both ($K \times 2, M$) and ($2 \times K, M$) X channels with complex constant channel gains.  This result shows that the layered interference alignment can almost surely characterize the DoF of these channels.

The proof of this extended theorem is provided in Appendix \ref{complex-case}. This theorem (see \ref{kgt:com}) shows that the total achievable DoF of MIMO X channel with complex channel gains  will be twice that of a similar channel with constant real gains. This observation can be justified either by relying on the fact that for a  complex channel, two dimensions per transmission (real and imaginary) are required, or by using the modified definition of DoF for complex channel gains and complex signal transmission, which is
\begin{equation}
d^i=\lim_{P \rightarrow \infty} \frac{R^i}{log{P}}.
\end{equation}

\section{conclusion}
In this paper, we introduced a new interference management tool, named  layered interference alignment.  We proved several metrical theorems in the field of Diophantine approximation which empowers using joint processing and simultaneous decoding. It is observed that, unlike GIC and SISO X channel, joint processing is required to characterize the total DoF of  MIMO X channels. To this end, we incorporated both the vector and the real interference alignment techniques for signal transmission, and relied on joint processing for simultaneous decoding. The total DoF of ($K \times 2,M$) and ($2 \times K,M$) X channels are characterized. It is observed that, for both complex and real channel gains, these can achieve the DOF upper bound of $\frac{2KM}{K+1}$.
\appendix
%\section*{Appendix}
%\renewcommand{\thesubsection}{\Alph{subsection}}
We start off with an introduction to the classical metric Diophantine approximation: the branch of number theory which can roughly be described as answering a simple question concerning `how well a real number can be approximated by rationals'. In subsequent subsections, we prove the Khintchine--Groshev type theorems for the particular type of linear forms, that are needed for the layered interference alignment.

\subsection{Khintchine--Groshev theorem for linear forms}\label{SKG}
In what follows, by an \emph{approximating function} we mean a decreasing function $\psi:\mb R^+\to \mb R^+$ such that $\psi(r)\to 0$ as $r\to\infty$.  An $m \times n$
matrix $\mbf{X} = (x_{i,j}) \in \mb{R}^{mn}$ is said to be
$\psi$-approximable if the system of inequalities
\begin{equation}\label{sysa}\|q_1x_{1,i}+q_2x_{2,i}+\ldots+q_mx_{m,i}\|<\psi{(|\mbf{q}|)} \ \ \ \left(1\leq i\leq n \right)\end{equation}

\noindent is satisfied for infinitely many vectors
$\mbf{q}\in\mb{Z}^m\setminus \{\mbf{0}\}$. Here $\|\cdot\|$ means
distance to the nearest integer. For clarity, equation \eqref{sysa} may be expressed in the form
\begin{equation}\label{sys1}
|q_1x_{1,i}+q_2x_{2,i}+\ldots+q_mx_{m,i}-p_i|<\psi{(|\mbf{q}|)} \ \ \ \left(1\leq i\leq n \right)
\end{equation}
\noindent which is satisfied for infinitely many vectors $(\mbf{p, q})=(p_1,\cdots, p_n, q_1,\cdots, q_m)\in\mb Z^n\times\mb{Z}^m\setminus \{\mbf{0}\}$.

The system
\[q_1x_{1,i}+q_2x_{2,i}+\ldots+q_mx_{m,i} \ \ \ \left(1\leq i\leq n \right)\]
of $n$ real linear forms in $m$ variables $q_1, \ldots, q_m$ will be
written more concisely as $\mbf{qX}$, where the matrix $\mbf X=
(x_{i,j})% := \left(
%              \begin{array}{ccc}
%                x_{1,1} & \ldots & x_{1,n} \\
%                x_{2,1} & \ldots &x_{2,n} \\
%                \vdots &  & \vdots \\
%                 x_{m,1} & \ldots &x_{m,n}
%              \end{array}
%            \right)
 $ is regarded as a point in $\mb{R}^{mn}$. It is easily seen
that $\psi$-approximability is unaffected under translation by
integer vectors, and we can therefore restrict attention to the  unit
cube $\mb{I}^{mn}$  as $$ \mb{R}
^{mn}\hm=\underset {\mbf{K}\in \mb{Z}^{mn}}{{\bigcup }}\left(
\mb{I}^{mn}+\mbf{K}\right).$$

The $\psi$-approximability in the
linear forms setup takes its roots from the linear form version of the
Dirichlet's theorem.

\begin{Dirichlet2}\label{dir}
  Let $N$ be a given natural number and let $\mbf X \in
  \mb{I}^{mn}$. Then there exists a non-zero integer  $\mbf{q} \in \mb{Z}^m$
  with $1 \leq |\mbf{q}|\leq N$ satisfying the system of inequalities
  \begin{equation*}
    \|q_1x_{1,i}+q_2x_{2,i}+\ldots+q_mx_{m,i}\|<N^{-\frac{m}{n}} \ \ \ \left(1\leq i\leq n \right).
  \end{equation*}

\end{Dirichlet2}

%\noindent It can be easily deduced from  Dirichlet's theorem that

\begin{corollary}For any $\mbf X \in   \mb{I}^{mn}$ there exist infinitely many integer vectors $\mbf{q} \in \mb{Z}^m$
  such that
  \begin{equation}\label{2.1}
  \|q_1x_{1,i}+q_2x_{2,i}+\ldots+q_mx_{m,i}\|<|\mbf{q}|^{-\frac{m}{n}} \ \ \ \left(1\leq i\leq n
  \right).
  \end{equation}
\end{corollary}

\noindent The right-hand side of (\ref{2.1}) may be sharpened by a
constant $c(m,n)$, but the best permissible values for $c(m,n)$ are
unknown except for $c(1, 1)=1/\sqrt 5$. %It leads naturally to the notion of

\noindent \emph{\textbf{Notation.}}  To simplify notation in the proofs below the Vinogradov symbols $\ll$ and $\gg$ will be used to indicate
an inequality with an unspecified positive multiplicative constant.
If $a\ll b$ and $a\gg b$  we write $a\asymp
b$, and say that the quantities $a$ and $b$ are comparable. Throughout, for any set $A$,  $|A|_l$ denote the $l$--dimensional Lebesgue measure of the set $A$.

The main result in the linear form settings is the
Khintchine--Groshev theorem, which gives an elegant answer to the
question of the size of the set $W(m,n;\psi)$. %The result links
%the measure of the set to the convergence or otherwise of a series
%that depends only on the approximating function and is the template
%for many results in the field of metric number theory. It provides a
%complete answer to the question of Lebesgue measure of the
%$\psi$-approximable points.
The following statement is due to
Groshev \cite{Groshev} and extends Khintchine's simultaneous result
\cite{K1} to the dual form case.

\begin{kgt} \label{KG} Let $\psi$ be an approximating function. Then

  $$
| W\left( m,n;\psi \right) |_{mn}\ =  \left \{
\begin{array}{cl}
0& {\rm \ if}  \qquad
 \sum \limits_{r=1}^{\infty}r^{m-1}\psi^n (r)<\infty, \, \\
1 & { \rm \ if}  \qquad  \sum\limits_{r=1}^{\infty} \
 r^{m-1}\psi^n (r)=\infty.\,
\end{array}
\right.
$$
\end{kgt}

\noindent The proof of the convergence case of the  Khintchine--Groshev
theorem is easily established by a straightforward application of
the Borel--Cantelli lemma and is free from any assumption on $\psi$.
The divergence part constitutes the main substance of the
theorem and requires the monotonicity assumption
on the function $\psi$. For further details and overview of this result we we refer the reader to \cite{BDV_mtl, mhty} and references therein.

\subsection{A mixed type Diophantine approximation}\label{new-problem}

In this section we  provide a set of new tools for decoder design of layered interference alignment. The tools needed  for the simple decoder design should empower the possibility of simultaneously decoding each part of each message in all antennas of each receiver. In the other words, considering limiting ourself to transmit integer numbers, we need to find the best estimator function that can estimate different linear forms of rational basis simultaneously.

Let $\psi$ be an approximating function and let $W_A (m,n;
\psi)$ be the set of $\mbf X\in\mb{I}^{mn}:=[-1/2, 1/2]^{mn}$ obtained by fixing the vector $(p_1, \cdots, p_n)$ in \eqref{sys1} as $(p, \cdots, p)$, i.e., the system of equations
\begin{equation}\label{eq1}
  |q_1x_{1,i}+q_2x_{2,i}+\ldots+q_mx_{m,i} -p|<\psi(|\mbf q|)\ \ \ \ 1\leq i\leq n
\end{equation}
 is satisfied for infinitely many $(p, \cdots, p, q_1, \cdots, q_m)\in\mb{Z}^n\times\mb{Z}^m\setminus\{\mbf{0}\}.$

The set $W_A(m,n;\psi)$ is  a hybrid of the classical set in which the distance to the nearest integer is allowed to vary from one linear form to the other. In the current situation it is the same for all the linear forms. Sets of similar nature has been studied by Hussain and his collaborators in \cite{DHMT, HK, mhjl2}. We prove the Khintchine--Groshev type result for $W_A (m,n;\psi)$. The results  crucially depend upon the choices of $m$ and $n$, similar to the above-mentioned papers and unlike the classical sets.

\begin{theorem}\label{thm1}
  Let $m+1>n$ and $\psi $
be an approximating function; then
\[
|W_A\left( m,n;\psi \right)|_{mn} = \left\{
\begin{array}{cl}
0& {\rm \ if}  \ \
 \sum \limits_{r=1}^{\infty}\psi^n (r)r^{m-n}<\infty \, \\
1  & { \rm \ if}  \ \  \sum\limits_{r=1}^{\infty} \ \psi^n
(r)r^{m-n}=\infty.\,
\end{array}
\right.
\]
\end{theorem}
The convergence half follows from the Borel--Cantelli lemma by construction of a suitable cover for the set $W_A\left( m,n;\psi \right)$. It does not rely on the choices of $m$ and $n$, and it is free from monotonic assumption on the approximating function. It is worth pointing out that for the application purposes the convergence case is all that matters. By setting $\mbf p=0$ in the above setup, a similar application of the convergence half already exists in achieving MIMO capacity within a constant gap \cite{OrdUri}. In fact, a particular form of Theorem \ref{kgt:com1} can be used to obtain the complex version of the results obtained in \cite{OrdUri}.  Fischler et al \cite{FHKL} has also used the convergence analogue of the above theorem for $\mbf p=0$ and for multiple approximating functions in proving the converse to linear independence criterion for several linear forms. It is worth demonstrating that for $\psi(r)=r^{{-\frac{m+1}{n}+1}-\epsilon}, \epsilon>0$,
\begin{eqnarray*}
  \sum \limits_{r=1}^{\infty}\psi (r)^{n}r^{m-n}=\sum \limits_{r=1}^{\infty}r^{-1-\epsilon}&<&\infty\quad {\rm if} \ \epsilon>0\\ &=& \infty\quad {\rm if} \ \epsilon\leq 0.
\end{eqnarray*}
From here it should be clear (if not, see Remark $1$ below) that for $m+1>n$ and $\epsilon>0 $, the set
\[|\{X\in W_A\left( m,n;\psi \right): d_{min}(X, R)\leq R^{{-\frac{m+1}{n}+1}-\epsilon} \ {\rm for \ i.m.} \ R\in \mb N\}|_{mn}=0.\]

%Now we formally prove the convergence case.
%The divergence case can be proved by using the similar arguments as in \cite{DHMT}.

\subsubsection{Proof of Theorem \ref{thm1}: the Convergence Case}\label{thm1-proof}

 Define the resonant sets as
\begin{equation*}
R_{q}=\left\{ \mbf X\in \mb I^{mn}:\mbf {qX}-\mbf p=0\right\}.
\end{equation*}

\noindent Thus, the resonant sets are $(m-1)n$-dimensional hyperplanes
passing through the point $\mbf p$. The set $W_A(m,n;\psi)$ can be written as a $\limsup$ set
using the resonant sets in the following way.

\begin{equation*}
W_A\left( m,n;\psi \right) =\underset{N=1}{\overset{\infty }{%
{\bigcap }}}\underset{r>N}{{\bigcup }}\underset{R_q:\left\vert \mbf{q}%
\right\vert =r}{{\bigcup }}B\left( R_{q},\psi (\vert\mbf{q}%
\vert )\right)
\end{equation*}

\noindent where
\begin{equation*}
B\left( R_{q},\psi (\vert\mbf{q}%
\vert )\right) =\left\{ \mbf X\in \mb I^{mn}: \text{dist}\left(
X,R_{q}\right) \leq \frac{\psi (\left\vert \mbf{q}\right\vert )}{%
\left\vert \mbf{q}\right\vert }\right\}.
\end{equation*}

\begin{figure}
\begin{center}
\begin{tikzpicture}[scale=0.75]
 %\begin{axis}[hide axis]
\draw (-2.5,3.5) rectangle(-1.1,2.5);
 \draw (-1.9,2.7)rectangle(-0.5,1.5);
 \draw (-1.3,1.7) rectangle(0.1,0.5);
\draw (-0.7,0.7) rectangle(.7,-0.5);

\draw (-0.1,-0.3) rectangle(1.3,-1.5);
 \draw (0.5,-1.3)rectangle(1.9,-2.5);
 \draw (1.1,-2.3) rectangle(2.5,-3.5);
  \draw (1.7,-3.3) rectangle(3.5,-4.5);

\draw (-5.5,3.5)%node[above left] {$(-\frac{1}{2},\frac{1}{2})$}
rectangle(3.5,-4.5)%node[below right]
%{$(\frac{1}{2},-\frac{1}{2})$}
node[above right] {$\mb I^2$};
 %\draw (4,3.5) node [black,above]
%{$(\frac{1}{2},\frac{1}{2})$};

%\draw (-4,-3.5) node [black,below] {$(-\frac{1}{2},-\frac{1}{2})$};
 \draw (-2.5,4.5) -- (2.5,-4.5);
 \draw[<-] (-2.2,4)--(-3.2,4.8) node[above] {$
 R_{\mbf{q}}$};
  \draw[<-] (2.3,-4.5)--(3.2,-5.5) node[right] {$ B(R_{\mbf{q}}, \psi(|\mbf{q}|))$};
  \draw[dashed,thick] (-2.2,3.5) -- (2.2,-4.5);
\draw[dashed, thick](-1.7,3.5) -- (2.8,-4.5); \draw[->] (-1,-1) --
(5.5,3.5)node[above] {$\mbf{q}=(q_1,q_2)$};
 \draw[->] (-5.75,-1) -- (5.5,-1)  coordinate(x axis)node [below] {$x$};
 \draw[->] (-1,-5.5) -- (-1,5.5)  coordinate(y axis)node [left] {$y$};

\draw [yshift=-0.05cm,below]
% (0,0) node [gray] {$(0, 0)$}
 (3.7,-0.250) node [black,below right] {$(\frac{1}{2},0)$};

\draw [yshift=-0.05cm,below] (-5.5,-0.25) node [black,below left]
{$(-\frac{1}{2},0)$};
\draw[fill=black](-1, 1.8) circle (0.04)node[left, xshift=-1.1cm]{$\tiny \frac{p}{q_2}$};
\draw[->, bend left] (-2.2, 1.8) to (-1.1, 1.8);

\draw[fill=black](0.55, -1) circle (0.04)node[left, xshift=-.7cm, yshift=-0.7cm]{$\tiny \frac{p}{q_1}$};
\draw[->, bend right] (0.5, -1.05) to (-0.35, -1.4);

%\node at (){$\tiny \frac{p}{q_2}$};

\draw [yshift=-0.05cm,above] (-0.9,3.5) node [black,above right]
{$(0,\frac{1}{2})$};

\draw [yshift=-0.05cm,below] (-0.9,-4.5) node [black,below right]
{$(0,-\frac{1}{2})$};
\end{tikzpicture}
\caption{The resonant set $R_{\mbf{q}}$ is a line for $m=2$ and $n=1$. The resonant
set $R_{\mbf{q}}$ is a line $q_1x+q_2y-p=0$, intercepting the $x$ and $y$ axes at
 $\frac{p}{q_1}$ and $\frac{p}{q_2}$,  respectively. The set $B\left( R_{\mbf{q}},\psi (\vert\mbf{q}%
 \vert )\right)$ is the $\frac{\psi (|\mbf{q}|)}{|\mbf q|}$ neighbourhood of $R_{\mbf{q}}$.}\label{khafan}
 \end{center}
\end{figure}
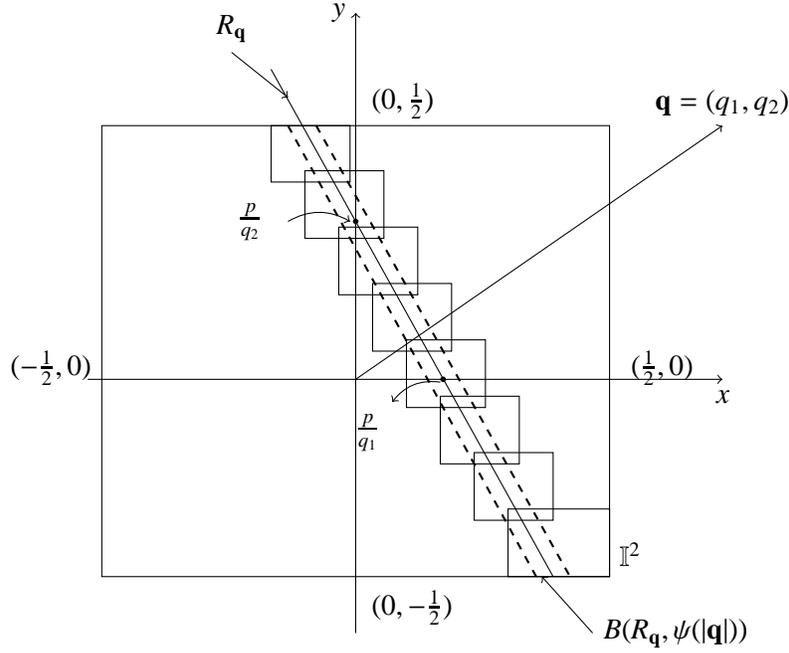

\medskip

%\noindent

Thus, for each $N\in \mb{N}$ the family
$
\left\{ \underset{R_{q}:\left\vert \mbf{q}\right\vert =r}{\bigcup }B\left( R_{q},\psi (\vert\mbf{q}%
\vert )\right) :r=N,N+1,...\right\}
$ is a cover for the set $W_A\left( m,n;\psi \right)$.
 Now, for each resonant set $R_{\mbf{q}},$ let $\Delta (\mbf{q})$ be a collection of $mn$-%
dimensional closed hypercubes $C$ with disjoint interiors and side
length comparable with $\psi (|\mbf q|)/|\mbf q|$ and
diameter at most $\psi (|\mbf q|)/|\mbf q|$ such that $C  \ \cap   \  \underset{R_{\mbf{q}}:\left\vert \mbf{q}\right\vert =r}{{\bigcup }}%
B\left( R_{\mbf{q}},\psi (\left\vert \mbf{q}\right\vert)\right)
\neq \emptyset$ and $ B\left( R_{\mbf{q}},\psi (\left\vert \mbf{q}\right\vert)\right) \subset
\underset{C\in \Delta (\mbf{q})}{{\bigcup }}C.$

\noindent Then%
\begin{equation*}
\# \Delta(\mbf{q})\ll \left( \psi (|\mbf q|)/|\mbf q|\right) ^{-(m-1)n}
\end{equation*}
\noindent where $\#$ denotes cardinality. Note that
\begin{eqnarray*}
W_A\left( m,n;\psi \right) &\subset &\underset{r>N}{{\bigcup
}}\underset{R_q:\left\vert \mbf{q}\right\vert =r}{{\bigcup
}}\Delta\left(
R_{\mbf{q}},\Psi (\left\vert \mbf{q}\right\vert)\right)
\subset \underset{r>N}{{\bigcup }} \ \underset{\Delta
(\mbf{q}):\left\vert \mbf{q}\right\vert =r}{{\bigcup }} \
\underset{C\in \Delta (\mbf{q})}{{\bigcup }}C.
\end{eqnarray*}

\noindent Hence,
\begin{eqnarray*}
\left\vert W_A\left( m,n;\psi \right) \right\vert_{mn} &\leq &
\sum\limits_{r>N}\sum\limits_{\Delta
(\mbf{q}):\left\vert \mbf{q}\right\vert =r} \sum\limits_{C\in \Delta (\mbf{q})}
|C|_{mn} \\
&\ll & \sum \limits_{r>N} r^{m}\left( \frac{\psi (r)}{r}\right)
^{mn}\left( \frac{\psi (r)}{r}\right) ^{-(m-1)n} \\
&=&\sum \limits_{r>N} r^{m-n}\psi (r)^{n}.
\end{eqnarray*}

\noindent Since the sum $\sum \limits_{r=1}^{\infty}\psi (r)^{n}r^{m-n}$
is convergent, which gives zero Lebesgue measure by the Borel--Cantelli
lemma.

\subsubsection{Proof of Theorem \ref{thm1}: the Divergence Case}

For the divergence case the ubiquity theorem
 \cite[Theorem 1]{bvubiq} is used, and to establish ubiquity two technical lemmas (Lemma \ref{d} and Lemma \ref{ubiquitymixed}) are needed. The work is similar to \cite{DHMT}; therefore, we only prove one of them and refer the interested reader to the aforementioned article \cite{DHMT}. Most of the metric results (Khintchine--Groshev, Jarnik, Jarnik--Besicovitch, and Schmidt theorems) stem from the Dirichlet type result which is stated and proved below for the current settings. Throughout, we set $N=\{2^t: t\in\mathbb N\}$.

\begin{lemma}\label{d}
 For $N_0< N$, for each $\mbf X\in \mb{I}^{mn}$
 there exists a non-zero integer vector
$\mbf{q}$ in $ \mb{Z}^{m}$ and $\mbf p\in \mb Z^n$ with $|\mbf{q}|,
|\mbf p| \leq N $ for $N_0 $ large enough such that
\begin{equation*}
|\mbf {qX}-\mbf p| < (m+2)2N^{-\frac{m+1}{n}+1}.
\end{equation*}
\end{lemma}

\noindent \emph{Proof of Lemma} $\ref{d}$. For $|\mbf p|<N$  and those $\mbf{q}$ with non-negative components, there are $\left(N+1\right) ^{m}N$ possible vectors of the form
$\mbf {qX}-p$ for which
\[-\frac{m+2}{2}N\leq \mbf {qX}-\mbf p\leq \frac{m+2}{2}N.\]

Divide the cube with centre $\mbf{0}$ and side length
$(m+2)N$ in $\mb{R} ^{n}$ into $N^{m+1}$ smaller cubes
of volume $(m+2)^{n}N ^{n-m-1}$ and side
length $(m+2)N^{1-\frac{m+1}{n}}$. Since $N^m<\left( N+1\right) ^{m}$, there are at least two vectors $%
\mbf{q}_{1}\mbf X-\mbf p_1 \mbf{,q}_{2}\mbf X-\mbf p_2$, say, in one small cube. Therefore%
\begin{equation*}
\left\vert
(\mbf{q}_{1}\mbf{-q}_{2})\mbf X-(\mbf p_1-\mbf p_2)\right\vert <
(m+2)2N^{-\frac{m+1}{n}+1}.
\end{equation*}

Evidently $\mbf{q}_1-\mbf{q}_2\in\mb{Z}^m$ and
$|\mbf{q}_1-\mbf{q}_2|\leq N$. Also, $\mbf p_1-\mbf p_2\in \mb Z$ and
$|\mbf p_1-\mbf p_2|\leq~N$ by choices of $\mbf p_1$ and
$\mbf p_2$.

\begin{lemma}\label{ubiquitymixed}
The family $\mcal{R}_q:=\{R_q:\mbf q \in\mb Z^m\setminus\{\mbf 0\}\}$ is locally ubiquitous with respect to the
function $\rho:~\mb{N}~\rightarrow \mb{R} ^{+}$ where
$$\rho(t)=(m+2)2N^{-\frac{m+1}{n}+1}\omega(t)$$
\end{lemma}
\noindent and $\omega(t)$ is a positive real increasing function such that $\omega(t)\to\infty$ as $t\to\infty$. However, it is not very restrictive in the sense that it can always be assumed as a step function and hence does not appear in the sum condition; for details see \cite[page 83]{DHMT}.

\medskip

%\noindent
In view of Lemma \ref{d}, it is natural to  consider the
following badly approximable set.  Let ${\rm Bad} (m,n)$  denote
the set of   $\mbf X\in \mb{I}^{mn}$ for which  there exists a constant
$C(\mbf X)>0$ such that
\begin{equation*}\label{eqb}
|\mbf {qX}-\mbf p|>C(\mbf X)|\mbf{q}|^{-\frac{m+1}{n}+1}\;\;\;\text{for \ all}\;\;(\mbf p, \mbf{q})\in\mb{Z}^{m+n}.
\end{equation*}

More generally, from the convergence part of Theorem \ref{thm1}, it is then clear  that for almost every $X\in\mb{I}^{mn}$  there exists a constant
$C(\mbf X)>0$ such that \begin{equation}
 |\mbf {qX}-\mbf p|\geq c(X)\psi(|\mbf q|)\quad \text{for \ all}\;\;(\mbf p, \mbf{q})\in\mb{Z}^{m+n}\setminus\{\mbf 0\}
\end{equation}
and denote the set of all such numbers as ${\rm Bad} (c, m,n)$ and $\cup_{c>0}{\rm Bad} (c, m,n)=\mb{I}^{mn}\setminus W_A\left( m,n;\psi \right)$. Now since $|W_A\left( m,n;\psi \right)|_{mn}=0$ which implies that $|\cup_{c>0}{\rm Bad} (c, m,n)|_{mn}=1.$ The question of finding the Hausdorff dimension and measure of  each ${\rm Bad} (c, m,n)$ is not dealt here and we leave it for another sitting. However, for the set ${\rm Bad} (m,n)$ it is straightforward to establish the following result.

\begin{theorem}\label{thmbad}Let $m+1>n$; then
\[\dim \textbf {\rm Bad}(m,n)=mn\]
and for $m+1\leq n$
\[|\textbf {\rm Bad}(m,n)|_{mn}=1.\]
\end{theorem}

Proof of Theorem \ref{thmbad} follows from \cite{mh, hkbad} by setting $u=1$ in those papers. Now, for $m+1>n$,  since $\textbf {\rm Bad}(m,n)\subseteq \mb{I}^{mn}\setminus W_A(m, n;\psi)$, therefore $|\textbf {\rm Bad}(m,n)|_{mn}=0$.
\begin{rem}
It should be clear from Theorem \ref{thmbad} that the minimum distance between $\mbf q \mbf X$ and the nearest integer vector $(p,\cdots,p)$ is at least $C(\mbf X)|\mbf{q}|^{-\frac{m+1}{n}+1}$, where $C(\mbf X)>0$ is a constant. Loosely speaking, ${\rm Bad} (m,n)$ consists of all those points that stay clear of $(m-1)n$-dimensional hyperplanes having diameters proportional to  $|\mbf{q}|^{-\frac{m+1}{n}+1}$ centered at the hyperplanes $R_q$. Note that if the exponent ${-\frac{m+1}{n}+1}$  is replaced by ${-\frac{m+1}{n}+1}-\epsilon$ for $\epsilon>0$, then  the set ${\rm Bad} (m,n)$ is of full Lebesgue measure. It is very pleasing and aligned with our applications.

\end{rem}
\begin{rem}
In the case $m+1\leq n$, the set $W_A(m, n;\psi)$ is over determined  and lies in a subset of strictly lower dimension than $mn$.
 To see this, consider the case $m=n$ and $\det \mbf X\neq 0$. This would imply that the defining inequalities \eqref{eq1} take the form
 \[|\mbf q-\mbf p \mbf X^{-1}|\leq C(\mbf X)\psi(|\mbf q|),\] \noindent which is obviously not true for sufficiently large $\mbf q$.

  The same logic extends to all other cases. For each $m\times n$ matrix $\mbf X\in\mb R^{mn}$ with column
 vectors $\mbf x^{(1)},\dots,\mbf x^{(n)}$ define $\tilde{\mbf X}$ to be the
 $m\times (n-1)$ matrix with column vectors
 $\mbf x^{(2)},\dots,\mbf x^{(n)}$. The set $\Gamma\subset\mb R^{mn}$ is the
 set of $\mbf X\in\mb R^{mn}$ such that the determinant of each $m\times m$
 minor of $\tilde{\mbf X}$ is zero.

 Then it can be easily  proved that $W_A(m, n;\psi)\subset\Gamma$ when $m+1\le n$, which will lead to further investigations of  metric theory for the cases $m+1\leq n$. However, this is not within the scope of the present paper. Therefore, we will not address it any further and refer the interested reader to \cite{ DHMT, mhjl2}, which comprehensively discusses such cases.
%\begin{lemma}\label{dimlemma}
%For $m+1\le n$ the set $W_A(m, n;\psi)$ is contained in $\Gamma$,
%and $\dim \Gamma =(m-1)n+m<mn$. Thus,
%\[\dim W_A(m, n;\psi)\le (m-1)n+m\]
%\end{lemma}

\end{rem}

\vspace*{1ex}
\subsection{Metric Diophantine Approximation over Complex Numbers: Classical Setup}\label{complex-case} Most of the complex Diophantine approximation theory is analogous to what we have discussed in the previous sections.  Surprisingly, analogues of Khintchine--Groshev theorems for systems of linear forms for complex numbers is not proved todate. We  prove them here alongwith the analogous results for mixed type linear forms. To keep the acquisition compact  and the length of the paper under control, we  state only the important changes.

In the 19th century, Hermite and Hurwitz studied the approximation of complex numbers by the ratios of Gaussian integers, a natural analogue of approximation of real numbers by rationals, \[\mb \mbf {Z}[i]=\{p_1+ip_2\in\mb C: p_1, p_2\in\mb \mbf {Z}\}.\]  However, complex Diophantine approximation appears
to be more difficult than the real case. For example, continued fractions, so simple and
effective for real numbers, are not so straightforward for complex numbers. In other words,  the best possible analogue of Dirichlet's theorem cannot be derived by means of a continued fraction expansion approach.

We will discuss the problem for the linear form setup and will list the recent developments so far for the particular cases. Let $\Psi$ be an approximating function satisfying step function i.e. $\psi(r)=\psi([r])$, where $[r]$ is the integer part of $r$. An $m \times n$
matrix $\mbf {Z} =
(z_{i,j}) %:= \left(
%              \begin{array}{ccc}
%                z_{1,1} & \ldots & z_{1,n} \\
%                z_{2,1} & \ldots &z_{2,n} \\
%                \vdots &  & \vdots \\
%                 z_{m,1} & \ldots &z_{m,n}
%              \end{array}
%            \right)
\in \mb{C}^{mn}
 $
  is said to be $\Psi$-approximable if the system of inequalities
\begin{equation}\label{sys}|q_1z_{1,j}+q_2z_{2,j}+\cdots+q_mz_{m,j}-p_j|<\Psi{(|\mbf{q}|_2)} \ \ \ \left(1\leq j\leq n \right)\end{equation}

\noindent is satisfied for infinitely many vectors
$\mbf{p\times q}\in\mb{Z}^n[i]\times \mb{Z}^m[i]\setminus \{\mbf{0}\}$. Throughout, the system \eqref{sys} will be written more concisely as $\mbf q\mbf {Z}$. Here $|\mbf q|_2=\max\{|q_1|_2, \cdots, |q_m|_2\}$, where for $q_k=q_{k_1}+iq_{k_2}\in \mb Z[i]$,  $|q_k|_2=\sqrt{|q_{k_1}|^2+|q_{k_2}|^2}$.

As in the real case, the stemming point of such approximation properties is the Dirichlet theorem. A short and more direct geometry of numbers
proof of the complex version of Dirichlet's theorem is given below. Although the constant
here is not best possible, the result is all that is needed to prove the complex analogue
of Khintchine--Groshev and Schmidt  type theorems without recourse to the hyperbolic space framework.

\begin{theorem}\label{dir:com} Given any $\mbf {Z}\in \mb C^{mn}$ and $N\in\mb N$, there exist Gaussian integers $\mbf p\in \mb{Z}^n[i]$ and non-zero $\mbf q\in \mb{Z}^m[i]$ with $0<|\mbf q|_2\leq N$ such that

\begin{equation}\label{sys2}|\mbf q \mbf {Z}-\mbf p|< \frac{c}{N^{m/n}} \end{equation} where $c>0$ is an appropriate constant. Moreover, there are infinitely many $(\mbf {p, q})\in \mb{Z}^n[i]\times \mb{Z}^m[i]\setminus \{\mbf{0}\}$ such that
\begin{equation*}\label{sys3}|\mbf q \mbf {Z}-\mbf p|< \frac{c}{|\mbf  q|_2^{m/n}}.\end{equation*}

\end{theorem}

\noindent\emph{Proof of Theorem } $\ref{dir:com}$. For clarity we prove the theorem for $m=2, n=1$. The proof of the case $m=n=1$ can be found in \cite{Dod_Kri05}. Let $\mbf {Z}=(x_1+iy_1, x_2+iy_2), \mbf q=(q_{1,1}+iq_{1,2}, q_{2,1}+iq_{2,2})$, and $p=(p_1+ip_2)$. Then
\begin{equation*}
  |\mbf q \mbf {Z}-\mbf p|=|q_{1,1}x_1+q_{2,1}x_2-q_{1,2}y_1-q_{2,2}y_2-p_1+i(q_{1,2}x_1+q_{2,2}x_2+q_{1,1}y_1+q_{2,1}y_2-p_2)|.
\end{equation*}

Consider the convex body
\begin{equation*}
  B=\left\{(q_{1,1}, q_{1,2}, q_{2,1}, q_{2,2}, p_1, p_2):\max\{q_{1,1}^2+ q_{1,2}^2, q_{2,1}^2+ q_{2,2}^2\}\leq N^2, \Delta\leq R^2\right\}
\end{equation*}
where
\begin{equation*}
  \Delta= \left(q_{1,1} x_1+q_{2,1} x_2 - q_{1,2} y_1 - q_{2,2} y_2 - p_1\right)^2+\left(q_{1,2} x_1 + q_{2,2} x_2 + q_{1,1} y_1 + q_{2,1} y_2 - p_2\right)^2.
\end{equation*}
Then
\begin{eqnarray*}
  |B|&=&\int_{\max\{q_{1,1}^2 + q_{1,2}^2 , \ q_{2,1}^2 + q_{2,2}^2\}\leq N^2}\int_{\Delta\leq R^2}dq_{1,1} dq_{1,2} dq_{2,1} dq_{2,2} dp_1 dp_2 \\ &=& \int_{\max\{q_{1,1}^2 + q_{1,2}^2 , \ q_{2,1}^2 + q_{2,2}^2\}\leq N^2}\pi R^2 dq_{1,1} dq_{1,2} dq_{2,1} dq_{2,2}\\&=& \pi^3 R^2 N^4\geq 2^6,
\end{eqnarray*}
if $R>\frac{2^3}{\pi^{3/4} N^2}.$ Hence, by Minkowski's theorem \cite{Har_Wri}, equation \eqref{sys2} has a non-zero integer solution with $0<|\mbf q|_2\leq N$.
\medskip

This result should be compared with the real Dirichlet's theorem in \S \ref{dir} for $m=4, n=2$. The complex points for which Theorem \ref{dir:com} cannot be improved by an arbitrary constant are called badly approximable. That is, a point
$\mbf {Z}\in\mb{C}^{mn}$ is said to be badly approximable if there
exists a constant $C(\mbf {Z})>0$ such that
\[|\mbf{q}\mbf {Z}-\mbf p|>C(\mbf {Z})|\mbf{q}|_2^{-\frac{m}{n}}\]
for all $(\mbf{p,  q})\in\mb{Z}^n[i]\times \mb{Z}^m[i]$.  Let
$\textbf{Bad}_{\mb C}(m,n)$ denote the set of  badly approximable
points in $\mb{C}^{mn}$.

The Hausdorff dimension of the set $\textbf{Bad}_{\mb C}(1,1)$ has been studied by various authors in different frameworks;  see, for instance, \cite[\S 5.3]{KTV}  in which authors determined the Hausdorff dimension for $\textbf{Bad}_{\mb C}(1,n)$, i.e.,
\[\dim \textbf{Bad}_{\mb C}(1,n)=n.\]
In fact, as a consequence of the general framework in their paper, they proved the Hausdorff dimension to be maximal in the weighted analogue of $\textbf{Bad}_{\mb C}$  intersected with any compact subset of $\mb C^n$. There framework can not be applied for the dual setup at work. However, it is reasonable to suspect that the Hausdorff dimension for $\textbf{Bad}_{\mb C}(m,n)$ is maximal or more generally for any compact subset $K\subset\mb C^{mn}$,
\begin{conj}
  $\dim \textbf{Bad}_{\mb C}(m,n)\cap K=\dim K$
\end{conj}
The treatment required to deal with this problem needs delicate number theoretic tools which would put this paper out of focus. Therefore, we will not deal with it any further.

\medskip
From now onwards we restrict ourselves to the $mn$-dimensional unit disc $D:=\left(\mb{C}\cap\Omega\right)^{mn}$, where $\Omega=\{a+ib: 0\leq a, b<1\}$, instead of considering the full space $\mb C^{mn}$.  The reason behind this restriction is that it is convenient to work in the unit discs, and the approximable properties (both well and bad) are invariant under the translation by the Gaussian integers. Let $W_\mb C(m,n;\Psi)$ denote the set of $\Psi$-approximable points in $D$ i.e.
\begin{equation*}\label{2.2}
W_\mb C(m,n;\Psi):=\left\{ \mbf {Z}\in D:|\mbf q \mbf {Z}-\mbf p|<\Psi (|\mbf{q}|_2 )\text{\ for i.m.\ } \ (\mbf {p, q})\in \mb{Z}^n[i]\times \mb{Z}^m[i]\setminus \{\mbf{0}\}\right\}.
\end{equation*}

\subsubsection{Khintchine--Groshev Theorem for complex numbers}

The aim here is to prove the complex version of the Khintchine--Groshev theorem

\begin{theorem}\label{kgt:com}Let $\Psi$ be an approximating function. Then

  $$
| W_\mb C\left( m,n;\Psi \right) |_{mn}\ =  \left \{
\begin{array}{cl}
0& {\rm \ if}  \qquad\sum \limits_{r=1}^{\infty}r^{2m-1}\Psi^{2n} (r)<\infty \, \\
Full & { \rm \ if}  \qquad  \sum_{r=1}^{\infty} \ r^{2m-1}\Psi^{2n} (r)=\infty.\,
\end{array}
\right.
$$
\end{theorem}

\noindent Here $|W_\mb C\left( m,n;\Psi \right)|_{mn}$ denotes the complex $mn$-dimensional Lebesgue measure of the set $W_\mb C\left( m,n;\Psi \right).$
For $m=n=1$, Theorem \ref{kgt:com}  was proved in $1952$ by LeVeque \cite{Leveque}, who combined Khintchine's continued
fraction approach with ideas from hyperbolic geometry. In $1982$, Sullivan \cite{Sul82} used Bianchi groups and some powerful
hyperbolic geometry arguments to prove more general Khintchine theorems for real and for
complex numbers. In the latter case, the result includes approximation of complex numbers
by ratios $p/q$ of integers $p, q$ from the imaginary quadratic fields $\mb R(i\sqrt d)$, where $d$ is a
square-free natural number. The case $d = 1$ corresponds to the Picard group and approximation
by Gaussian rationals. The result was also  derived by Beresnevich et al. as a consequence of  ubiquity framework in \cite[Theorem 7]{BDV_mtl}.

\subsubsection{Proof of the Convergence Case of Theorem \ref{kgt:com}} \label{kgt:com:proof}

As before,  Theorem \ref{kgt:com} is proved for the case $m=2, n=1$, leaving behind the obvious modifications to deal with the higher dimensions. First, the convergence case is dealt with. The resonant set is defined as
\begin{eqnarray*}
  C_{\mbf q}&:=&\left\{\mbf {Z}\in D:|\mbf q \mbf {Z}-\mbf p|=0\right\}\\ &=& \left\{(x_1+iy_1, x_2+iy_2)\in D:|(q_{1,1} + iq_{1,2}, q_{2,1} + iq_{2,2})\cdot(x_1+iy_1, x_2+iy_2)-(p_1+ip_2)|=0\right\}\\&=& \left\{(x_1+iy_1, x_2+iy_2)\in D: \begin{array}{l}
  q_{1,1} x_1+q_{2,1} x_2-q_{1,2} y_1-q_{2,2} y_2=p_1   \ \ \text{and} \\
  q_{1,2} x_1+q_{2,2} x_2+q_{1,1} y_1+q_{2,1} y_2=p_2
                                  \end{array}\right\}.
\end{eqnarray*}

The set $W_\mb C\left( 2,1;\Psi \right)$ can be written  using the resonant sets
\begin{equation*}
W_\mb C\left( 2,1;\Psi \right) =\underset{N=1}{\overset{\infty }{%
{\bigcap }}}\underset{r>N}{{\bigcup }}\underset{C_{\mbf q}:|\mbf p|_2<| \mbf{q}|_2 =r }{{\bigcup }}B\left( C_{\mbf q},\Psi (|\mbf{q}|_2 )\right)
\end{equation*}

\noindent where
\begin{equation*}
B\left( C_{\mbf q},\Psi (|\mbf{q}|_2 )\right) =\left\{ \mbf {Z}\in D: \text{dist}\left(
\mbf {Z},C_{\mbf q}\right) \leq \frac{\Psi (|\mbf{q}|_2 )}{%
| \mbf{q}|_2 }\right\}.
\end{equation*}
It follows that $$W_\mb C\left( 2,1;\Psi \right)\subseteq \underset{r>N}{{\bigcup }}\underset{C_{\mbf q}:|\mbf p|_2<| \mbf{q}|_2 =r }{{\bigcup }}B\left( C_{\mbf q},\Psi (|\mbf{q}|_2 )\right).$$  In other words, $W_\mb C\left( 2,1;\Psi \right)$ has a natural cover $\mcal C=\left\{B\left( C_{\mbf q},\Psi (|\mbf{q}|_2 )\right) :| \mbf{q}|_2>N\right\}$ for each $N=1, 2, \cdots$. It can further be covered by a collection of $4$-dimensional hypercubes with disjoint interior and side length comparable with $\Psi (|\mbf{q}|_2)/|\mbf{q}|_2$. The number of such hypercubes is clearly $\ll\left(\Psi (|\mbf{q}|_2)/|\mbf{q}|_2\right)^{-2}$. Thus,
\begin{eqnarray}\label{count}
 \left\vert W_\mb C\left( 2,1;\Psi \right)\right\vert_2 &\leq& \sum_{r=N}^\infty\sum_{C_{\mbf q}:|\mbf p|_2<| \mbf{q}|_2 =r }^\infty \left\vert B\left( C_{\mbf q},\Psi (|\mbf{q}|_2 )\right)\right\vert_2\notag
 \\ &\ll&  \sum_{r=N}^\infty\sum_{r<| \mbf{q}|_2 <r+1 }\left(\Psi (|\mbf{q}|_2)/|\mbf{q}|_2\right)^{-2} \left(\Psi (|\mbf{q}|_2)/|\mbf{q}|_2\right)^4\notag
  \\ &=& \sum_{r=N}^\infty \left(\Psi (r)/r\right)^2\sum_{r<| \mbf{q}|_2 <r+1 }1.
\end{eqnarray}
Now it remains to count $\sum\limits_{r<| \mbf{q}|_2 <r+1 }1$. An argument from \cite[p. 328]{Dod_Kri05} or \cite[Th. 386]{Har_Wri} is followed to conclude that $\sum_{r<| \mbf{q}|_2 <r+1 }1\ll r^5$. Thus, \eqref{count} becomes

\begin{equation*}
  \left\vert W_\mb C\left( 2,1;\Psi \right)\right\vert_2\ll \sum_{r=N}^\infty r^3\Psi (r)^2.
\end{equation*}
Now, since the sum $\sum_{r=N}^\infty r^3\Psi (r)^2<\infty$, the tail of the series can be made arbitrarily small. Hence, by the Borel--Cantelli lemma, $\left\vert W_\mb C\left( 2,1;\Psi \right)\right\vert_2=0$.

The divergence case of the above theorem can be similarly proved by following the similar arguments as in the real case. Precisely, one would need to utilize the ubiquity framework to extend \cite[Th. 7]{BDV_mtl} for the linear forms setup. The Dirichlet theorem \ref{dir:com} would again be used to prove the ubiquity lemma.  The details are left for the interested reader.

\subsubsection{A Complex Hybrid Setup}

As in the previous section, let $\Psi$ be an approximating function satisfying the step function. An $m \times n$
matrix $\mbf {Z} \in \mb{C}^{mn}$
  is said to be $\Psi$-approximable if the system of inequalities
\begin{equation}\label{sysch}|q_1z_{1,j}+q_2z_{2,j}+\cdots+q_mz_{m,j}-p|<\Psi{(|\mbf{q}|_2)} \ \ \ \left(1\leq j\leq n \right)\end{equation}

\noindent is satisfied for infinitely many vectors $(p, \cdots, p, q_1, \cdots, q_m)\in\mb{Z}^n[i]\times \mb{Z}^m[i]\setminus \{\mbf{0}\}$. That is, the system \eqref{sysch} is obtained by keeping the nearest integer vector $(p, \cdots, p)$  the same for all the linear forms. Since the results are very similar to $W_A(m, n;\Psi)$ and can be proved analogously, they are only stated here with obvious modifications. The first one is the Dirichlet type theorem, and rest of the results stem from it. It also serves the purpose of finding the minimum distance between $\mathbf q \mbf {Z}$ and $\mathbf p$.
\begin{theorem}\label{dir:com1} Given any $\mbf {Z}\in \mb C^{mn}$ and $N\in\mb N$, there exist Gaussian integers $\mbf p=(p_1+i p_2, \cdots, p_1+i p_2)\in \mb{Z}^n[i]$ and non-zero $\mbf q=(q_{11}+i q_{12}, \cdots, q_{m1}+i q_{m2})\in \mb{Z}^m[i]$ with $0<|\mbf q|_2\leq N$ such that

\begin{equation*}\label{sys4}|\mbf q \mbf {Z}-\mbf p|< \frac{c}{N^{\frac{m+1}{n}-1}} \end{equation*} where $c>0$ is an appropriate constant. Moreover, there are infinitely many $(\mbf {p, q})\in \mb{Z}^n[i]\times \mb{Z}^m[i]\setminus \{\mbf{0}\}$ such that
\begin{equation*}\label{sys5}|\mbf q \mbf {Z}-\mbf p|< c|\mbf  q|_2^{-\frac{m+1}{n}+1}.\end{equation*}

\end{theorem}
Let $W_{\mb C_A}(m,n;\Psi)$ denote the set of $\Psi$-approximable points in $D$, i.e., the set of points that satisfy the system \eqref{sysch}. Then, one has the analogue of the Khintchine--Groshev theorem for this setup.
\begin{theorem}\label{kgt:com1}Let $\Psi$ be an approximating function and let $m+1>n$. Then

  $$
| W_{\mb C_A}\left( m,n;\Psi \right) |_{mn}\ =  \left \{
\begin{array}{cl}
0& {\rm \ if}  \qquad \sum \limits_{r=1}^{\infty}\left(r^{m-n}\Psi^n (r)\right)^2<\infty \, \\
Full & { \rm \ if}  \qquad  \sum_{r=1}^{\infty} \left(r^{m-n}\Psi^n (r)\right)^2=\infty.\,
\end{array}
\right.
$$
\end{theorem}

The proof of this theorem is again similar to that of Theorem \ref{kgt:com}. The details are left for the  interested reader.
%\newpage

\def\cprime{$'$}

\end{document}